\documentclass[11pt]{article}
\usepackage{latexsym}

\newcommand{\newsection}{ \setcounter{equation}{0} \section}

\newcommand{\beq}{\begin{equation}} \newcommand{\eeq}{\end{equation}}
\newcommand{\bea}{\begin{eqnarray}} \newcommand{\eea}{\end{eqnarray}}

  \newcommand
{\Romannumeral}[1]{\uppercase\expandafter{\romannumeral#1}}

\newcommand{\be}{\begin{enumerate}} \newcommand{\ee}{\end{enumerate}}
\newcommand{\bi}{\begin{itemize}} \newcommand{\ei}{\end{itemize}}
\newcommand{\ba}{\begin{array}} \newcommand{\ea}{\end{array}}
\newcommand{\bc}{\begin{center}} \newcommand{\ec}{\end{center}}
\newcommand{\bt}{\begin{tabular}} \newcommand{\et}{\end{tabular}}

\def\lsim{\mathrel{\rlap{\lower4pt\hbox{\hskip1pt$\sim$}}
    \raise1pt\hbox{$<$}}}           % less than or approx. symbol
\def\gsim{\mathrel{\rlap{\lower4pt\hbox{\hskip1pt$\sim$}}
    \raise1pt\hbox{$>$}}}           % greater than or approx. symbol

\newcommand{\half}{\textstyle {1\over2} \displaystyle}    % One half
\newcommand{\third}{\textstyle {1\over3} \displaystyle}   % One third
\newcommand{\Dslash}{{\hbox{D}\kern-0.6em\raise0.15ex\hbox{/}}} % D slash

\begin{document}

\setlength{\oddsidemargin}{0cm} \setlength{\baselineskip}{7mm}

\input epsf

\begin{normalsize}\begin{flushright}
%    UCI-2011-xx \\
August 2011 \\
\end{flushright}\end{normalsize}

\begin{center}
  
\vspace{10pt}

% {\Large \bf Gravitational Slip Function}
% {\Large \bf with a Scale-Dependent Newton's Constant G }

{\Large \bf Scale-Dependent Newton's Constant G }

{\Large \bf in the Conformal Newtonian Gauge}

\vspace{20pt}

{\sl Herbert W. Hamber}
$^{}$\footnote{HHamber@uci.edu} \\
% Department of Physics and Astronomy \\
% University of California \\
% Irvine, CA 92697-4575, USA \\
Institut des Hautes Etudes Scientifiques \\
% I.H.E.S. \\
35, route de Chartres \\
91440 Bures-sur-Yvette, France. \\

\vspace{5pt}

and

{\sl Reiko Toriumi}
$^{}$\footnote{RToriumi@uci.edu}
\\
Department of Physics and Astronomy \\
University of California \\
Irvine, California 92697-4575, USA \\

\vspace{10pt}

\end{center}

\begin{center} {\bf ABSTRACT } \end{center}

\noindent
In classical gravity deviations from the predictions of the Einstein theory are often discussed
within the framework of the conformal Newtonian gauge, where scalar perturbations are 
described by two potentials $\phi$ and $\psi$.
In this paper we use the above gauge to explore possible cosmological consequences of a running Newton's constant $ G ( \Box ) $,
as suggested by the nontrivial ultraviolet fixed point scenario arising from the quantum field-theoretic
treatment of Einstein gravity with a cosmological constant term.
Here we  focus on the effects of a scale-dependent coupling 
on the so-called gravitational slip functions $\eta = \psi / \phi -1 $, whose 
classical general relativity value is zero.
Starting from a set of manifestly covariant but nonlocal effective field equations
derived earlier, we compute the leading corrections in the potentials $\phi$ and $\psi$ for a
nonrelativistic, pressureless fluid.
After providing an estimate for the quantity $\eta$, we then focus on a comparison 
with results obtained in a previous paper on matter density perturbations in the synchronous gauge,
which gave an estimate for the growth index parameter $\gamma$, also in the presence of a 
running $G$.
Our results indicate that, in the present framework and for a given $ G ( \Box ) $, the corrections tend
to be significantly larger in magnitude for the perturbation growth exponents than 
for the conformal Newtonian gauge slip function.

% \vspace{15pt}
% \begin{center} {\it (Submitted to the Physical Review D)} \end{center}
% PACS codes are 04.60.-m, 04.60.Gw and 98.80.Qc 

\vfill

%% no page number on 1-st page

\pagestyle{empty}

\newpage

\pagestyle{plain}

\vskip 10pt
\newsection{Introduction}
\hspace*{\parindent}
\label{sec:intro}

Recent years have seen the development of a fascinating variety of alternative theories
of gravity, in addition to the more traditional alternate frameworks, which used to include
just Brans-Dicke, higher derivative, effective quantum gravity and supergravity theories.
Some of the new additions to the by now rather long list include dilaton gravity, 
$f(R)$ gravity, torsion gravity, loop quantum gravity, holographic modified gravity, 
and a few others, just to cite here a few representative examples. 
All of these theories eventually predict some level of deviation from classical gravity,
at short- or long-distance scales, 
which is often parametrized either by a suitable set of post-Newtonian parameters, or
more recently, by the introduction of a gravitational slip function \cite{dam06,dam93}.

In this paper, we will focus on the analysis of departures from general relativity (GR) in the 
gravitational slip function, obtained in the framework of the conformal Newtonian gauge,
and within the rather narrow context of the nontrivial ultraviolet fixed point scenario 
for Einstein gravity with a cosmological term.
Thus, instead of looking at deviations from GR at very short distances, due
to new interactions such as the ones suggested by string theories \cite{ven02},
we will be considering here infrared effects, which could manifest themselves 
at very large distances.

The specific nature of the scenario we will be investigating here is motivated
by the field-theoretic treatment of models for quantum gravity, based on the 
(minimal) Einstein action with a bare cosmological term.
The theory's long-distance scaling properties used as the basis for the present
work follow from the existence of a nontrivial ultraviolet 
fixed point of the renormalization group in Newton's constant $G$.
The latter is inaccessible by direct perturbation theory in four dimensions, 
and can be shown to radically alter the short- and long-distance behavior
of the theory when compared to more naive, perturbative expectations.
The renormalization- group origin of such fixed points was first discussed in detail by 
Wilson some time ago for scalar and self-coupled fermion theories \cite{wil72}.
The general field-theoretic methods were later extended and
applied to gravity, where they are now referred to as 
the nontrivial UV fixed point, or asymptotic safety, scenario \cite{wei77}.
It is fair to say that so far this is the only field-theoretic approach known 
to work consistently in other not perturbatively renormalizable theories, 
such as the nonlinear sigma model above two dimensions \cite{zin02}.
While perhaps still a bit mundane in the context of gravity, such nontrivial
fixed points are well studied and well understood in statistical field theory,
where they generally describe phase transitions between ordered and
disordered ground states, or between weakly coupled and condensed states.

The paper is organized as follows.
First (Sec. 2) we recall the effective covariant field equations describing the running
of $G$, and describe briefly the nature of various objects and parameters 
entering the quantum nonlocal corrections;
a more complete description of the basic setup can be found in our previous 
papers on the subject, and will not be repeated here.
We then discuss the zeroth order (in the metric fluctuations) field equations
and energy-momentum conservation equations
for the standard homogeneous isotropic metric, with a running $G (\Box)$.
Later (Sec. 3) we extend the formalism to deal with small metric and matter
perturbations, and list the relevant field and energy conservation
equations to first order in the perturbations in the comoving gauge.
These above results are then (Sec. 4) reexpressed in two other choices of gauge, 
the synchronous and the conformal Newtonian gauge.
The latter choice of gauge allows us to extract an expression for the
gravitational slip function $\eta$ due to $G(\Box)$ (Sec. 5).
This quantity is then evaluated within the context of a $\Lambda CDM $ model, 
for redshifts corresponding to the present era (z=0).
The resulting correction is then compared to current astrophysical
observations, as well as to our previous results (and observations) 
regarding the corrections due to $G(\Box)$ to the matter density 
perturbation growth exponents.
The conclusions provide an interpretation of the theoretical
results, and their associated uncertainties, in view of 
present and future high precision determination of the gravitational
slip function and growth exponents.

\vskip 40pt
\newsection{Running Newton's Constant ${\bf G(\Box)}$}
\hspace*{\parindent}
\label{sec:gbox}

As mentioned in the introduction,
it is not the purpose of this paper to provide a satisfactory description, or motivation,
for the running of $G$ that arises in the quantum-field-theoretic treatment of Einstein's
gravity with a cosmological term.
Here we only provide a brief summary, and only the most relevant formulas will
be given for later reference; 
a more complete set of references can be found, for example, in \cite{hbook}.

The running of Newton's constant $G$ has been computed both on the lattice in four dimensions
\cite{hw84,ham00}, and in the continuum within the framework of the background 
field expansion applied to $d=2+\epsilon$ spacetime dimensions \cite{wei77,eps},
and later also using truncation methods applied in $d=4$ \cite{reu98}.
In either case one obtains a momentum-dependent $G(k^2)$, which
eventually needs to be reexpressed in a suitable coordinate-independent way, 
so that it can be consistently applied to more general problems, involving
arbitrary background geometries.
The first step in analyzing the consequences of a running of $G$
is therefore to rewrite the expression for $G(k^2)$ in a coordinate-independent
way, either by the use of a nonlocal Vilkovisky-type effective gravity action \cite{vil84,ven90},
or by the use of a set of consistent effective field equations.
In going from momentum to position space one employs 
$k^2 \rightarrow - \Box$, which then gives for
the quantum-mechanical running of the gravitational
coupling the replacement $ G  \;\; \rightarrow \;\; G( \Box ) $.
Then the running of $G$  is given in the vicinity of the UV fixed point by
\beq
G ( \Box ) \, = \, G_0 \left [ \; 1 \, 
+ \, c_0 \left ( { 1 \over \xi^2 \Box  } \right )^{1 / 2 \nu} \, 
+ \, \dots \, \right ] \; ,
\label{eq:grun_box}
\eeq
where $\Box \equiv g^{\mu\nu} \nabla_\mu \nabla_\nu$
is the covariant d'Alembertian, and the dots represent higher
order terms in an expansion in $1 / ( \xi^2 \Box ) $.
Current evidence from Euclidean lattice quantum gravity points toward 
$c_0 > 0$ (implying infrared growth) and $\nu \simeq \third $ \cite{ham00}.
Within the quantum-field-theoretic renormalization-group treatment, the quantity $\xi$ arises 
as an integration constant of the Callan-Symanzik renormalization-group equations.

One issue of great relevance to the physical interpretation of the results, 
is therefore a correct identification of the renormalization-group invariant scale $\xi$.
A number of arguments, mostly based on nonperturbative lattice results and scaling
considerations involving the gravitational Wilson loop and its relevance for large scale
observable curvature \cite{loo07}, can be given in support of the suggestion 
that the dynamically generated infrared 
cutoff scale $\xi$ (analogous to the $\Lambda_{\overline{MS}}$ of QCD) 
can be quite large in the case of gravity (for a recent review, see Ref. \cite{hbook}).
These arguments would then suggest that the new scale $\xi$ is naturally expected to
be related to the large scale average curvature, and thus could be of cosmological relevance,
\begin{equation}
\lambda  \;  \simeq \;  { 3 \over \xi^2 } \; .
\label{eq:xi_lambda}
\end{equation} 
These considerations then lead to a more concrete quantitative estimate 
for the scale in the running $G(\Box)$ of Eq.~(\ref{eq:grun_box}), namely
$\xi \sim 1 / \sqrt{\lambda/3} \sim 1.51 \times 10^{28} {\rm cm} $.
Moreover, from these types of arguments one would also infer that the constant 
$G_0$ in Eq.~(\ref{eq:grun_box}) can, to a very close 
approximation, be identified with the laboratory value of 
Newton's constant, $ \sqrt{G_0} \sim 1.6 \times 10^{-33} {\rm cm}$.
The running of $G$ envisioned above would then remain in agreement with laboratory
and solar system precision tests of general relativity.

The appearance of the d'Alembertian $\Box$ in the running of $G$ naturally 
leads to both a nonlocal effective gravitational action and a corresponding
set of nonlocal modified field equations.
In the simplest scenario, instead of the ordinary Einstein field equations with constant $G$
\beq
R_{\mu\nu} \, - \, \half \, g_{\mu\nu} \, R \, + \, \lambda \, g_{\mu\nu}
\; = \; 8 \pi \, G \, T_{\mu\nu} \; ,
\label{eq:field}
\eeq
one is now led to consider the modified effective field equations
\beq
R_{\mu\nu} \, - \, \half \, g_{\mu\nu} \, R \, + \, \lambda \, g_{\mu\nu}
\; = \; 8 \pi \, G  ( \Box )  \, T_{\mu\nu} 
\label{eq:field1}
\eeq
with the nonlocal term due to the $G(\Box)$.
By being manifestly covariant, these equations still satisfy some
of the basic requirements for a set of consistent field equations
incorporating the running of $G$.
Not unexpectedly though, the new nonlocal equations are much
harder to solve than the original classical field equations for constant $G$.

The effective nonlocal field equations of 
Eq.~(\ref{eq:field1}) can be recast in a form very similar to the classical
field equations, but with a new source term 
$ {\tilde T_{\mu\nu}} \, = \, \left [ G  ( \Box )  / G_0   \right ] \, T_{\mu\nu}$
defined as the effective, or gravitationally dressed, energy-momentum tensor \cite{hw05,hw06}.
Ultimately the consistency of the effective field equations demands that it
be exactly conserved, in consideration of the contracted Bianchi identity 
satisfied by the Ricci tensor.
In this picture, therefore, the running of $G$ can be viewed 
as contributing to a sort of vacuum fluid, introduced in order to account for 
the new gravitational quantum vacuum-polarization contribution.

Due the appearance of a negative fractional exponent in Eq.~(\ref{eq:grun_box}),
the covariant operator appearing in the expression for $G(\Box)$
has to be suitably defined by analytic continuation. 
This can be done, for example, by computing $\Box^n$ for positive
integer $n$, and then analytically continuing to $n \rightarrow -1/2\nu$ \cite{hw05}.
Equivalently, $G(\Box)$ can be defined via a regulated parametric integral 
representation \cite{lop07}, such as
\beq
\left ( { 1 \over - \Box (g) + \mu^2 } \right )^{1/ 2 \nu } 
\, = \, 
{ 1 \over \Gamma ( { 1 \over 2 \nu } ) } \,
\int_0^\infty d \alpha \; \alpha^{  1 / 2 \nu - 1 } \;
e^{  - \alpha \, ( - \Box (g) + \mu^2 ) } \; ,
\label{eq:gbox_exp}
\eeq
where $\mu \rightarrow 0 $ is a suitable infrared regulator.
As far as the calculations in this paper are concerned, it will not be necessary
to commit oneself to an unduly specific form for the running of $G (\Box )$.
Thus, for example, although the lattice gravity results only allow for a nondegenerate phase
for the case $c_0 >0$, it will nevertheless be possible later to have either sign
for the correction in Eq.~(\ref{eq:grun_box}).
We note here that a running cosmological constant $\lambda (k) \rightarrow \lambda ( \Box ) $ 
causes a number of mathematical inconsistencies \cite{hw05,ht10} within
the manifestly covariant framework, described here
by the effective field equations of Eq.~(\ref{eq:field1}).
Indeed if one assumes for the running part of $ \lambda ( \Box ) \sim ( \xi^2 \Box )^{-\sigma} $,
 then the infrared regulated expression
in Eq.~(\ref{eq:gbox_exp}) gives no running of $\lambda$, after using the identity
$\nabla_\lambda g_{\mu\nu} =0 $.
\footnote{
To be a bit specific, consider the case of a scale dependent
$\lambda (k)$, which we will write here as $\lambda = \lambda_0 + \delta \lambda (k) $.
Let us also assume, for concreteness, that $\delta \lambda (k) \sim c_1 (k^2)^{-\sigma}$,
where $c_1$ and $\sigma$ are some constants, and then make the transition to coordinate 
space by using $ k^2 \rightarrow - \Box $.
Thus $\delta \lambda ( \Box ) \sim ( - \Box + \mu^2 )^{- \sigma} $, where one should
be careful and use the infrared regulated expression in Eq.~(\ref{eq:gbox_exp}).
The effective field equations will then contain a term 
$ {1 \over 2} \, \delta \lambda (\Box) \cdot g_{\mu\nu}  $
$ = {1 \over 2 } \, c_1 \; { 1 \over \Gamma ( { \sigma } ) } \,
\int_0^\infty d \alpha \; \alpha^{  \sigma - 1 } \;
e^{  - \alpha \, ( - \Box (g) + \mu^2 ) } \cdot g_{\mu\nu}  
$ $ = { 1 \over 2} \, c_1 \, ( \mu^2 )^{- \sigma} \cdot g_{\mu\nu} $,
which is still gives just a constant multiplying the metric $g_{\mu\nu}$.}
This last conclusion is in general agreement with the field-theoretic results of the 
nontrivial renormalization-group fixed point scenario \cite{hbook},thereby providing perhaps an independent consistency check.
Note that this rather general argument also applies to possible additional contributions 
from nonzero 
vacuum expectation values of matter fields, such as the Higgs.
As a result, in the present quantum-field-theoretic motivated framework $\lambda$ 
is assumed not to run.

\vskip 40pt
\subsection{ Zeroth order effective field equations with ${\bf G(\Box)}$}
\hspace*{\parindent}
\label{sec:field_gbox}

A scale-dependent Newton's constant is expected to lead to small modifications
of the standard cosmological solutions to the Einstein field equations.
Here we will summarize what modifications are
expected from the effective field equations on the basis of $G(\Box)$.
The starting point is the quantum effective nonlocal field equations
of Eq.~(\ref{eq:field1}), 
with $G(\Box)$ defined in Eq.~(\ref{eq:grun_box}).
In the Friedmann-Lema\^itre-Robertson-Walker (FLRW) framework these are
applied to the standard homogeneous isotropic metric
\beq
d \tau^2 \; = \;  dt^2 - a^2(t) \left \{ { dr^2 \over 1 - k\,r^2 } 
+ r^2 \, \left ( d\theta^2 + \sin^2 \theta \, d\varphi^2 \right )  \right \} \;\;\;\; k =0, \pm1 \; .
\label{eq:frw}
\eeq
In the following, we will only consider the case $k=0$ (spatially flat universe).
It should be noted that there are in fact {\it two} related quantum contributions to the
effective covariant field equations. 
The first one arises because of the presence of a nonvanishing 
cosmological constant $\lambda \simeq 3 / \xi^2 $, caused by the
nonperturbative quantum vacuum condensate $<R> \, \neq 0$ \cite{loo07}.
As in the case of the standard FLRW cosmology, this is expected to be 
the dominant contributions at large times $t$, and gives an exponential
(for $\lambda>0$), or cyclic (for $\lambda < 0$) expansion of the scale factor.
The second contribution arises because of the explicit running of $G (\Box)$ in the 
effective field equations.

The next step therefore is a systematic examination of the nature of
the solutions to the full effective field equations,
with $G ( \Box )$ involving the relevant covariant d'Alembertian operator
\beq
\Box \; = \; g^{\mu\nu} \, \nabla_\mu \nabla_\nu 
\label{eq:box}
\eeq
acting on second rank tensors as in the case of $T_{\mu\nu}$.
To start the process, we will assume that $T_{\mu\nu}$ is described by the perfect fluid form, 
\beq
T_{\mu \nu} = \left [ \, p(t) + \rho(t) \, \right ] u_\mu \, u_\nu + g_{\mu \nu} \, p(t)
\label{eq:tmunu_perf}
\eeq
for which one needs to compute the action of $\Box^n$ on $T_{\mu\nu}$,  and 
then analytically continues the answer to negative fractional values of $n = -1/2 \nu $.
The results of \cite{hw05,hw06,lop07,ht10}  then show  that a 
nonvanishing pressure contribution is generated in 
the effective field equations, even if one initially assumes a pressureless fluid, $p(t)=0$.
After a somewhat lengthy derivation one obtains for a universe filled with nonrelativistic 
matter ($p$=0) the following set of effective Friedmann equations,
\bea
{ k \over a^2 (t) } \, + \,
{ \dot{a}^2 (t) \over a^2 (t) }  
& = & { 8 \pi \, G(t) \over 3 } \, \rho (t) \, + \, { \lambda \over 3 }
\nonumber \\
& = & { 8 \pi \, G_0 \over 3 } \, \left [ \,
1 \, + \, c_t \, ( t / \xi )^{1 / \nu} \, + \, \dots \, \right ]  \, \rho (t)
\, + \, { \lambda \over 3 }
\label{eq:fried_tt}
\eea
for the $tt$ field equation, and
\bea
{ k \over a^2 (t) } \, + \, { \dot{a}^2 (t) \over a^2 (t) }
\, + \, { 2 \, \ddot{a}(t) \over a(t) } 
& = & - \, { 8 \pi \, G_0 \over 3 } \, \left [ \, c_t \, ( t / \xi )^{1 / \nu} 
\, + \, \dots \, \right ] \, \rho (t) 
\, + \, \lambda
\label{eq:fried_rr}
\eea
for the $rr$ field equation.
In the above expressions, the running of $G$ appropriate for 
the Robertson-Walker metric is
\beq
G (t) \, \equiv \, G_0 \left ( 1 + { \delta G(t) \over G_0 } \, \right ) 
\, = \, G_0 \left [ 1 + c_t \, 
\left ( { t \over \xi } \right )^{1 / \nu} \, + \, \dots \right ] 
\label{eq:grun_t}
\eeq
with $c_t$ of the same order as $c_0$ in Eq.~(\ref{eq:grun_k}) \cite{hw05}
(in the quoted reference the estimate $c_t \simeq 0.450 \; c_0$ was given for the tensor box operator).
From the above form of $\delta G(t)$ one sees that the amplitude of the quantum correction
is actually proportional  to the combination $c_0 / \xi^3$ for $\nu=1/3$.
Note also that the running of $G$ induces an effective pressure term in the second 
($rr$) equation, due to the presence of a relativistic fluid whose origin is in
the vacuum-polarization contribution.
Another noteworthy general feature of the new field equations is the additional power-law
acceleration contribution, on top of the standard one due to the $\lambda$ term.

\vskip 40pt
\subsection{Introduction of the ${\bf w_{vac}}$ parameter}
\hspace*{\parindent}
\label{sec:w_vac}

It was noted in \cite{hw05,ht10} that the field equations with a running $ G $, Eqs.~(\ref{eq:fried_tt})
and (\ref{eq:fried_rr}), can be recast in an equivalent, but slightly more appealing, 
form by defining a vacuum-polarization pressure $p_{vac}$ and density 
$\rho_{vac}$, such that for the FLRW background one has
\beq
\rho_{vac} (t) = {\delta G(t) \over G_0} \, \rho (t)  \;\;\;\;\;\;\;\;\;\;\;\; 
p_{vac} (t) = { 1 \over 3} \, {\delta G(t) \over G_0} \, \rho (t) \; .
\label{eq:rhovac_t}
\eeq
From this viewpoint, the inclusion of a vacuum-polarization contribution in the FLRW 
framework seems to amount to a replacement 
\beq
\rho(t) \rightarrow \rho(t) + \rho_{vac} (t)     
\;\;\;\;\;\;\;\;\;\;\;\;
p(t) \rightarrow p(t) +  p_{vac} (t)
\label{eq:vac}
\eeq
in the original field equations.
Just as one introduces the parameter $w$, describing the matter equation of state, 
\beq
p (t) = w \,  \rho(t)
\label{eq:w_def}
\eeq
with $ w=0 $ for nonrelativistic matter, one can do the same for the remaining  contribution
by setting
\beq
p_{vac} (t) = w_{vac} \; \rho_{vac} (t) \; .
\label{eq:wvac_def}
\eeq
We should remark here that the original calculations \cite{hw05}, and more recently
\cite{ht10} which included metric perturbations, also indicate that 
\beq
w_{vac}= \third
\label{eq:wvac}
\eeq
is obtained {\it generally} for the given class of $G(\Box)$ considered, and is 
not tied therefore to a specific choice of $\nu$, such as $\nu=\third$.

The previous, slightly more compact, notation allows one to rewrite the field 
equations for the FLRW background in an equivalent form, which we will describe next.
We note here that, when dealing with density perturbations, we will have to distinguish
the background, which will involve a background pressure ($\bar p$) and background
density ($\bar \rho$), from the corresponding perturbations which will be denoted here
by $\delta p$ and $\delta \rho$.
With this notation and for constant $G_0$,
 the FLRW field equations for the background are written as
\bea
3 \, {{\dot{a}}^2 (t) \over {a}^2 (t)} 
& = & 
8 \pi \, G_0  \, \bar{\rho} (t) + \lambda 
\nonumber \\
{{\dot{a}}^2 (t) \over {a}^2 (t)} + 2\, {\ddot{a} (t) \over a (t)} 
& = &
- 8 \pi \, G_0 \, {\bar p}  (t) + \lambda \; .
\label{eq:fried_0}
\eea
Then in the presence of a running $G(\Box)$,  and in accordance with the results 
of Eqs.~(\ref{eq:fried_tt}) and (\ref{eq:fried_rr}), 
the modified FLRW equations for the background read
\bea
3 \, {{\dot{a}}^2 (t) \over {a}^2 (t)} 
& = & 
8 \pi \, G_0 \left ( 1 + {\delta G(t) \over G_0} \right ) \, \bar{\rho}(t) + \lambda 
\nonumber \\
{{\dot{a}}^2 (t) \over {a}^2 (t)} + 2\, {\ddot{a} (t) \over a (t)} 
& = & 
- 8 \pi \, G_0 \, \left ( w + w_{vac}  {\delta G(t) \over G_0} \right )\, \bar{\rho} (t) + \lambda \; ,
\label{eq:fried_run}
\eea
using the definitions in Eqs.~(\ref{eq:w_def}) and (\ref{eq:wvac_def}),
here with $\bar{p}_{vac} (t) = w_{vac} \,  \bar{\rho}_{vac}(t)$.

Of course the procedure of defining a $\rho_{vac}$ and a $p_{vac}$ contribution,
arising from quantum vacuum-polarization effects, is not necessarily
restricted to the FLRW background metric case.
In general one can decompose the full source term in the effective nonlocal
field equations of Eq.~(\ref{eq:field1}), making use of
\beq
G(\Box) = G_0 \, \left ( 1 \, +  {\delta G(\Box) \over G_0} \right ) 
\;\;\;\;\;\;  {\rm with} \;\;\;\;\;
{\delta G(\Box) \over G_0} \equiv c_0 \left ( { 1 \over \xi^2 \Box } \right )^{1 / 2 \nu}  \; ,
\label{eq:grun_box_1}
\eeq
 as two contributions,
\beq
{ 1 \over G_0 } \, G(\Box) \, T_{\mu\nu}  \, = \, 
\left ( 1 + {\delta G(\Box) \over G_0}  \right ) \, T_{\mu\nu}  \, = \,
T_{\mu\nu}   +  T_{\mu\nu}^{vac} \; .
\label{eq:tmunu_vac}
\eeq
The latter involves the nonlocal part
\beq
T_{\mu\nu}^{vac} \, \equiv \,  {\delta G(\Box) \over G_0} \, T_{\mu\nu} \; .
\label{eq:tmunu_vac1}
\eeq
Consistency of the full nonlocal field equations requires that the sum
be conserved,
\beq
\nabla^\mu \left ( T_{\mu\nu}   +  T_{\mu\nu}^{vac}  \right ) = 0 \; .
\eeq
In general one cannot expect that the contribution $ T_{\mu\nu}^{vac} $
will always be expressible in the perfect fluid form of Eq.~(\ref{eq:tmunu_perf}), even if the
original $ T_{\mu\nu} $ for matter (or radiation) has such a form.
The former will in general contain, for example, nonvanishing shear stress contributions, 
even if they were originally absent in the matter part.

%\newpage

\vskip 40pt
\newsection{Relativistic treatment of matter density perturbations}
\hspace*{\parindent}
\label{sec:pert}

Besides the modified cosmic scale factor evolution just discussed, 
the running of $G(\Box)$, as given in Eq.~(\ref{eq:grun_box}),
also affects the nature of matter density perturbations on large scales.
In computing these effects, it is customary to introduce a perturbed metric of
the form
\beq
{d\tau}^2 = {dt}^2 - a^2 \left ( \delta_{ij} + h_{ij} \right ) dx^i dx^j \; ,
\label{eq:pert_metric}
\eeq
with $a(t)$ the unperturbed scale factor and $ h_{ij} ({\bf x},t)$ a small
metric perturbation, and $h_{00}=h_{i0}=0$ by choice of coordinates.
After decomposing the matter fields into background and fluctuation contribution, 
$\rho = \bar{\rho}+\delta \rho$, $p = \bar{p}+\delta p $, and ${\bf v} = \bar{\bf v}+\delta {\bf v}$, 
it is customary in these treatments to expand the density, pressure and metric
perturbations in spatial Fourier modes,
\bea
\delta \rho ({\bf x},t) & = & \delta \rho_{\bf q} (t) \, e^{i \, {\bf q} \, \cdot \, {\bf x}}
\;\;\;\;\;\;\;\;
\delta p ({\bf x},t) = \delta p_{\bf q} (t) \, e^{i \, {\bf q} \,\cdot \, {\bf x}}
\nonumber \\
\delta {\bf v} ({\bf x},t) & = & {\delta {\bf v}}_{\bf q} (t)  \, e^{i \, {\bf q} \, \cdot \, {\bf x}}
\;\;\;\;\;\;\;\;
h_{ij} ({\bf x},t) = h_{ {\bf q} \, ij} (t)\, e^{i \, {\bf q} \, \cdot \, {\bf x}} 
\label{eq:fourier}
\eea
with ${\bf q}$ the comoving wave number.
Once the Fourier coefficients have been determined, the original perturbations can
later be obtained from
\beq
\delta \rho ({\bf x},t) \, = \, \int { d^3 {\bf x} \over ( 2 \pi )^{3/2} }
\, e^{ - i \, {\bf q} \, \cdot \, {\bf x}} \, \delta \rho_{\bf q} (t) 
\eeq
and similarly for the other fluctuation components.
Then the field equations with a constant $G_0$ [Eq.~(\ref{eq:field})]
are given to zeroth order in the perturbations by
Eq.~(\ref{eq:fried_0}),  which fixes the three background fields 
$a(t)$, $\bar{\rho} (t)$, and $\bar{p} (t) $.
Note that in a comoving frame the four-velocity appearing in Eq.~(\ref{eq:tmunu_perf})
has components $ u^i = 1, \; u^0 = 0 $.
Without $G(\Box)$, to first order in the perturbations and in the limit 
${\bf q} \rightarrow 0$ the field equations give
\bea
{\dot{a} (t) \over a (t)}\, \dot{h} (t) & = & 8 \pi \, G_0  \, \bar{\rho} (t) \, \delta (t) 
\nonumber \\
\ddot{h} (t) + 3 \, {\dot{a} (t) \over a (t)}\, \dot{h} (t) 
& = & - \, 24 \pi \, G_0  \, w \, \bar{\rho}(t) \, \delta (t)
\eea
with the matter density contrast defined as $\delta (t) \equiv \delta \rho (t) / \bar{\rho} (t) $, 
$h(t) \equiv h_{ii} (t)$ the trace part of $h_{ij}$, and $w=0$ for nonrelativistic matter.
When combined together, these last two equations then yield 
a single equation for the trace of the metric perturbation,
\beq
\ddot{h} (t) + 2 \, {\dot{a} (t) \over a (t)} \, \dot{h} (t) \; = \; 
 - \, 8 \pi \, G_0 ( 1 + 3\, w ) \, \bar{\rho}(t) \, \delta (t) \; .
\eeq
From first order energy conservation, one has 
$ - {1 \over 2}\, \left ( 1 + w \right )\, h (t) = \delta (t) $, 
which then allows one to eliminate $h(t)$ in favor of $\delta(t)$, which then
allows one to obtain a single second order equation for the density contrast $\delta(t)$.
In the case of a running $G(\Box)$, the above equations need to be rederived
from the effective covariant field equations of Eq.~(\ref{eq:field1}), and lead to several
additional terms not present at the classical level \cite{ht10}.

\vskip 40pt
\subsection{Zeroth order energy-momentum conservation}
\hspace*{\parindent}
\label{sec:enmom_zeroth}

As a first step in computing the effects of density matter perturbations, one
needs to examine the consequences of energy and momentum conservation,
to zeroth and first order in the relevant perturbations.
If one takes the covariant divergence of the field equations in Eq.~(\ref{eq:field1}),
the left-hand side has to vanish identically because of the Bianchi identity. 
The right-hand side then gives 
$\nabla^\mu \left (  T_{\mu\nu} + T_{\mu\nu}^{vac} \right ) =0 $,
where the fields in $ T_{\mu \nu}^{vac} $ can be expressed, at least to lowest order,
in terms of the $p_{vac}$ and $\rho_{vac}$ fields defined in Eqs.~(\ref{eq:rhovac_t}) and (\ref{eq:wvac_def}).
The first equation one obtains is the zeroth (in the fluctuations) order energy conservation
in the presence of $G(\Box)$, which reads
\beq
3 \, {\dot{a} (t) \over a (t)} \, 
\left [ \left (1+w \right ) + \left (1+ w_{vac} \right )\,{\delta G(t) \over G_0} 
\right ]  \bar{\rho} (t) 
 +   { \dot{\delta G}(t) \over G_0} \, \bar{\rho} (t)
 + \left ( 1 + {\delta G(t) \over G_0} \right )\, \dot{\bar{\rho}} (t)  = 0 \; .
\label{eq:encons_zeroth}
\eeq
In the absence of a running $G$ these equations reduce to the ordinary mass
conservation equation for $w=0$,
\beq
\dot{\bar{\rho}}(t) = - 3 \, { \dot{a}(t) \over a(t)} \, \bar{\rho}(t) \; .
\label{eq:encons_frw}
\eeq
It is often convenient to solve the energy conservation
equation not for ${\bar{\rho}} (t)$, but instead for ${\bar{\rho}} (a)$.
This requires that, instead of using the expression for $G(t)$ in Eq.~(\ref{eq:grun_t}),
one uses the equivalent expression for $G(a)$
\beq
G (a) = G_0 \left ( 1 + {\delta G (a) \over G_0}  \right ) \; ,
\label{eq:grun_a}
\eeq
which is easily obtained once the relationship between $t$ and $a(t)$ is known (see discussion later).
Note for example that the solution to Eq.~(\ref{eq:encons_zeroth}) can be written as
\beq
\bar{\rho} (a)= {\rm const. } \; \exp \left \{
- \int  { d a \over a} \;
\left (  3 + { \delta G (a) \over G_0 } + a \, { \delta G' (a) \over G_0  } \right ) \,
\right \} \; .
\label{eq:rho_zeroth_sim}
\eeq

\vskip 40pt
\subsection{Effective energy-momentum tensor involving ${\bf \rho_{vac}}$ and ${\bf p_{vac}}$}
\hspace*{\parindent}
\label{sec:enmom_vac}

The next step consists in obtaining the equations which govern the effects of 
small field perturbations.
These equations will involve, apart from the metric perturbation $h_{ij}$, the matter 
and vacuum-polarization contributions.
The latter arise from [see Eq.~(\ref{eq:tmunu_vac})]
\beq
\left ( 1 + {\delta G(\Box) \over G_0}  \right ) \, T_{\mu\nu}  \, = \,
T_{\mu\nu}   +  T_{\mu\nu}^{vac} 
% \label{eq:tmunu_vac}
\eeq
with a nonlocal 
$ T_{\mu\nu}^{vac} \equiv ( \delta G(\Box) / G_0 ) \, T_{\mu\nu} $.
Fortunately to zeroth order in the fluctuations the results of Ref. \cite{hw05} indicated 
that the modifications from the nonlocal vacuum-polarization term could 
simply be accounted for by the substitution
\beq
\bar{\rho} (t) \rightarrow \; \bar{\rho} (t) + {\bar\rho}_{vac} (t)     
\;\;\;\;\;\;\;\;\;\;\;\;
\bar{p} (t) \rightarrow \; \bar{p} (t) +  {\bar p}_{vac} (t) \; .
\label{eq:sub0}
\eeq
Here we will apply this last result to the small field fluctuations as well, and set
\beq 
\delta \rho_{\bf q} (t) \rightarrow \; \delta \rho_{\bf q} (t) + \delta \rho_{{\bf q} \, vac} (t)
\;\;\;\;\;\;\;\;\;\;\;\;
\delta p_{\bf q}  (t) \rightarrow \; \delta p_{\bf q}  (t) +  \delta p_{{\bf q} \, vac} (t) \; .
\label{eq:sub1}
\eeq
The underlying assumption is of course that the equation of state for the vacuum fluid still
remains roughly correct when a small perturbation is added.
Furthermore, just like we had $ {\bar p} (t) = w \,  \bar{\rho} (t) $ [Eq.~(\ref{eq:w_def})]
and  $\bar{p}_{vac} (t) = w_{vac} \,  \bar{\rho}_{vac}(t) $ [Eq. ~(\ref{eq:wvac_def})]
with $ w_{vac} = \third $, we now write for the fluctuations
\beq
\delta p_{\bf q}  (t) = w \, \delta \rho_{\bf q}   (t) \;\;\;\;\;\;\;\;\; 
\delta p_{{\bf q} \, vac} (t) = w_{vac} \, \delta \rho_{{\bf q} \, vac} (t) \; ,
\label{eq:wvac_fluc}
\eeq
at least to leading order in the long wavelength limit, ${\bf q} \rightarrow 0 $.
In this limit we then have simply
\beq
\delta p (t) = w \, \delta \rho   (t) \;\;\;\;\;\;\;\;\; 
\delta p_{vac} (t) = w_{vac} \, \delta \rho_{vac} (t) \equiv 
w_{vac}  \, {\delta G(t) \over G_0}  \delta \rho (t) \; ,
\label{eq:wvac_fluc1}
\eeq
with $G(t)$ given in Eq.~(\ref{eq:grun_t}), and we have used Eq.~(\ref{eq:rhovac_t}), 
now applied to the fluctuation $\delta \rho_{vac} (t)$,
\beq
\delta \rho_{vac} (t) \, = \, {\delta G(t) \over G_0}  \, \delta \rho (t) + \dots
\label{eq:delta_rhovac_t}
\eeq
where the dots indicate possible additional $O(h)$ contributions.
A bit of thought reveals that the above treatment is incomplete,
since $G(\Box)$ in the effective field equation of Eq.~(\ref{eq:field1}) 
contains, for the perturbed Robertson-Walker metric of Eq.~(\ref{eq:pert_metric}), 
terms of order $h_{ij}$,  which need to be accounted for in the effective $T^{\mu\nu}_{vac}$.
Consequently the covariant d'Alembertian operator 
$ \Box \; = \; g^{\mu\nu} \, \nabla_\mu \nabla_\nu  $
acting here on second rank tensors, such as $T_{\mu\nu}$,
\bea
\nabla_{\nu} T_{\alpha\beta} \, = \, \partial_\nu T_{\alpha\beta} 
- \Gamma_{\alpha\nu}^{\lambda} T_{\lambda\beta} 
- \Gamma_{\beta\nu}^{\lambda} T_{\alpha\lambda} \, \equiv \, I_{\nu\alpha\beta}
\nonumber
\eea
\beq 
\nabla_{\mu} \left ( \nabla_{\nu} T_{\alpha\beta} \right )
= \, \partial_\mu I_{\nu\alpha\beta} 
- \Gamma_{\nu\mu}^{\lambda} I_{\lambda\alpha\beta} 
- \Gamma_{\alpha\mu}^{\lambda} I_{\nu\lambda\beta} 
- \Gamma_{\beta\mu}^{\lambda} I_{\nu\alpha\lambda}  \; ,
\label{eq:box_on_tensors}
\eeq
needs to be Taylor expanded in the small field perturbation $h_{ij}$,
\beq
\Box (g) \, = \, \Box^{(0)}  + \Box^{(1)} (h) + O (h^2) \; .
\eeq
One then obtains for $G(\Box)$ itself
\beq
G(\Box) \, = \, G_0 \,  \left [ 
1 + \, { c_0 \over \xi^{1 / \nu} } \,  
\left ( { 1 \over \Box^{(0)} + \Box^{(1)} (h) + O(h^2) } \right )^{1/ 2 \nu}  + \dots
\right ]  \; ,
\label{eq:gbox_h}
\eeq
which requires the use of the binomial expansion for the operator
$ (A+B)^{-1} = A^{-1} - A^{-1} B A^{-1}  + \dots $.
Thus for sufficiently small perturbations it should be adequate to expand $G(\Box)$ 
entering the effective field equations in powers of the metric perturbation $h_{ij} $.
Next we turn to a discussion of the above results in different gauges.

\vskip 40pt
\newsection{Gauge choices and corresponding transformations}
\hspace*{\parindent}
\label{sec:gauge}

The previous discussion and summary focused exclusively on the comoving gauge
choice for the metric, implicit in the definition of Eq.~(\ref{eq:frw}).
Next we will consider some additional gauges.
In this paper we will specifically refer to {\it three} choices for the metric: the
comoving, synchronous and conformal Newtonian forms.
The first two are closely related to each other, and were used to obtain
part of the results presented in our previous work \cite{hw05,ht10}, 
which was summarized in the previous section.
Note that in our previous work \cite{ht10} we did not include the effects of
a stress field $s$, since it was not necessary for the discussion of density perturbations;
new terms arising from such a field are included below.
The third form of the metric is the primary focus of the present discussion;
the results obtained later on in this paper will either be derived for this metric,
or transformed to it by relying on results obtained previously in the other gauges.

\vskip 40pt
\subsection{Comoving, synchronous and conformal Newtonian gauges}
\hspace*{\parindent}
\label{sec:metric}

The {\it comoving} metric has the form
\beq
g_{\mu \, \nu} = \bar{g}_{\mu \, \nu} + h_{\mu \, \nu} \; ,
\eeq
with background metric
\beq
\bar{g}_{\mu \, \nu} = {\rm diag} \left(-1, a^2, a^2, a^2 \right) \; .
\eeq
For the fluctuation one sets
\beq
h_{0i} = h_{i0} = 0 \; ,
\eeq
and decomposes the remaining $h_{ij}$ as
\beq
h_{ij} ({\bf k},t) \; = \;  a^2
\left[ \, { 1 \over 3 }\,  h \, \delta_{ij} 
+ \left( {1 \over 3} \, \delta_{ij} - {k_i \, k_j \over k^2} \right) \, s  \right]
\label{eq:stress_def}
\eeq
so that $ Tr(h_{ij}) = a^2 \, h $.
Besides the scale factor $a$, the metric is therefore parametrized in terms of the
two functions $s$ and $h$.

On the other hand, in the {\it synchronous} gauge one sets again 
$ g_{\mu \, \nu} = \bar{g}_{\mu \, \nu} + h_{\mu \, \nu} $ 
now with background metric
\beq
\bar{g}_{\mu \, \nu} =a^2 \, {\rm diag} \left(-1, 1, 1, 1 \right) \; .
\eeq
For the fluctuation one sets again 
$ h_{0i} = h_{i0} = 0 $ and 
\beq
h_{ij} ({\bf k},t)
\; = \; a^2
\left[ \, {k_i \, k_j \over k^2} \, h_{sync} 
+ \left( {k_i \, k_j \over k^2} - {1 \over 3} \, \delta_{ij} \right) \, 6 \, \eta 
\right]  \; ,
\eeq
so that now $ Tr(h_{ij} ) = a^2 \, h_{sync} $.
Here, besides the overall scale factor $a$, the metric is parametrized in terms of the
two functions $\eta$ and $h_{sync}$.
From a comparison of the two gauges (comoving and synchronous) one has
\beq
2 \, \eta \; = \; - {1 \over 3} \left( h + s \right)
\eeq
and 
\beq
 h_{sync} + 6 \, \eta \; = \; - s  \; .
\eeq

Finally the {\it conformal Newtonian} gauge is in turn described by two scalar potentials 
$ \psi $ and $\phi $.
In this case the line element is given by
\beq
d\tau^2 = - g_{\mu \, \nu} dx^\mu \, dx^\nu = a^2 \, \Big\lbrace \left( 1 + 2 \, \psi \right) \, dt^2 - \left(1 - 2 \, \phi \right) \, dx_i \, dx^i \Big\rbrace \; .
\eeq
Therefore for the metric itself one writes again
$ g_{\mu \, \nu} = \bar{g}_{\mu \, \nu} + h_{\mu \, \nu} $
with
$ \bar{g}_{\mu \, \nu} =a^2 \, diag \left(-1, 1, 1, 1 \right) $
as for the synchronous case,  and furthermore 
$ h_{0i} = h_{i0} = 0 $ as before, and now
\beq
h_{00} = a^2 \, \left(- \, 2 \, \psi \right)
\eeq
\beq
h_{ij} = a^2 \, \left(- \, 2 \phi \right) \, \delta_{ij} \; .
\eeq
A suitable set of gauge transformations then allows one to go from the synchronous, or comoving, to the conformal Newtonian gauge \cite{mab95}.

\vskip 40pt
\subsection{Tensor box in the comoving gauge}
\hspace*{\parindent}
\label{sec:tensorbox}

To compute higher order contributions from the $h_{ij}$ 's appearing in the 
comoving  gauge metric, one needs to expand $G(\Box)$ in the various metric
perturbations,
\beq
G(\Box) = G_0 \, \left [ 
1 + \, { c_0 \over \xi^{1 / \nu} } \, 
\left (
\left ( { 1 \over \Box^{(0)} } \right )^{1 / 2 \nu} 
- {1 \over 2 \, \nu} \, { 1 \over \Box^{(0)} } \cdot \Box^{(1)} (h,s) \cdot
\left ( { 1 \over \Box^{(0)} } \right )^{1 / 2 \nu} \, + \dots
\right )
\right ] \; ,
\label{eq:gbox_hs_expanded}
\eeq
where the superscripts $(0)$ and $(1)$ refer to zeroth and first order in this expansion, respectively.
To get the correction of $O(h,s)$ to the field equations, one therefore needs to consider
the relevant term in the expansion of $ ( 1 + \delta G(\Box) / G_0 ) \, T_{\mu \nu} $,
\beq
- {1 \over 2 \, \nu} \, { 1 \over \Box^{(0)} } \cdot \Box^{(1)} (h,s) \cdot
{ \delta G ( \Box^{(0)} ) \over G_0 } \cdot T_{\mu \nu}
\; = \; 
- {1 \over 2 \, \nu} \, { c_0 \over \xi^{1 / \nu} } \,
{ 1 \over \Box^{(0)} } \cdot \Box^{(1)} (h,s) \cdot 
\left ( { 1 \over \Box^{(0)} } \right )^{1 / 2 \nu} \cdot T_{\mu \nu} \; .
\label{gbox_hs_tensor}
\eeq
This last form allows us to use the results obtained previously 
for the FLRW case, namely
\beq
{ \delta G ( \Box^{(0)} ) \over G_0 } \, T_{\mu \nu} \; = \; T_{\mu \nu}^{vac}
\eeq
with here
\beq
T_{\mu \nu}^{vac} \; = \;
\left [ p_{vac} (t) + \rho_{vac} (t) \right ] u_\mu \, u_\nu + g_{\mu \nu} \, p_{vac} (t)
\label{eq:tmunu_vac_2}
\eeq
to zeroth order in $h$, and
\beq
\rho_{vac} (t) = {\delta G(t) \over G_0} \, \bar{\rho} (t) \;\;\;\;\;\;\;\;\;\;\;\; 
p_{vac} (t) = w_{vac} \, {\delta G(t) \over G_0} \, \bar{\rho} (t) \; .
\label{eq:rhovac_t_1}
\eeq
and $w_{vac} = 1/3$.
Therefore, in light of the results of Ref. \cite{hw05}, the problem has been 
reduced to computing the more tractable expression
\beq
- {1 \over 2 \, \nu} \, 
{ 1 \over \Box^{(0)} } \cdot \Box^{(1)} (h,s) \cdot T_{\mu \nu}^{vac} \; .
\label{gbox_hs_tensor_vac}
\eeq
To make progress, we will assume a harmonic time dependence for both
the perturbations $h(t) = h_0 \, e^{i \omega t}$ and $s(t) = s_0 \, e^{i \omega t}$, and 
for the background quantities 
$a(t)=a_0 \, e^{i \Gamma t}$, $\rho (t)= \rho_0 \, e^{i \Gamma t}$, and \
$\delta G (t)= \delta G_0 \, e^{i \Gamma t}$.
From now on we shall consider both $\omega$ and $\Gamma$ as slowly varying
functions (indeed constants),
with the time scale of variations for the perturbation much shorter
than the time scale associated with all the background quantities.
A more sophisticated treatment will be reserved for future work.
Therefore we will take here $\omega \gg \Gamma$ or $ \dot{h} / h \gg \dot{a} / a $,
which is the same approximation that was used in obtaining the results of Ref. \cite{ht10}.

Let us now list, in sequence, the required matrix elements needed for the present calculation.
For the tensor box $tt$ matrix element 
$ ( - {1 \over 2 \, \nu} \, 
{ 1 \over \Box^{(0)} } \cdot \Box^{(1)} (h,s) \cdot T^{vac} )_{00} $ one obtains
\beq
+ \, {1 \over 2 \nu} \, {11 \over 3} \, {\delta G(t) \over G_0} \, \rho(t) \, {\Gamma \over \omega} \, h 
+ {\mathcal{O}} (k^2) \; .
\eeq
For the tensor box $ti$ matrix element 
$ ( - {1 \over 2 \, \nu} \, 
{ 1 \over \Box^{(0)} } \cdot \Box^{(1)} (h,s) \cdot T^{vac} )_{0i} $ one obtains
\beq
- \, i \, k_i \, {1 \over 2 \nu} \, {2 \over 9} \, {\delta G(t) \over G_0} \, \rho(t) 
\, {1 \over i \, \omega} \, \left( h - 2 \, s \right) + {\mathcal{O}} (k^2) \; .
\eeq
For the tensor box $ii$ matrix element, summed over $i$, 
$ ( - {1 \over 2 \, \nu} \, 
{ 1 \over \Box^{(0)} } \cdot \Box^{(1)} (h,s) \cdot T^{vac} )_{ii} $, one obtains
\beq
3 \, \left(+ \, {1 \over 2 \nu} \, w_{vac} \, { 11 \over 3 } \, a^2 \, {\delta G(t) \over G_0} \, \rho(t) \, {\Gamma \over \omega} \, h \right) + {\mathcal{O}} (k^2) \; .
\eeq
For the tensor box $ii$ matrix element, not summed over $i$, 
$ ( - {1 \over 2 \, \nu} \, 
{ 1 \over \Box^{(0)} } \cdot \Box^{(1)} (h,s) \cdot T^{vac} )_{ii} $, one obtains
\beq
+ \, {1 \over 2 \nu} \, a^2 \, {\delta G(t) \over G_0} \, \rho(t) 
\, \left[w_{vac} \, {11 \over 3} \, {\Gamma \over \omega} \, h 
+ {8 \over 9} \, \left(1 - 3 \, {k_i \over k^2} \, \right) \, {\Gamma \over \omega} \, s \right]
+ {\mathcal{O}} (k^2) \; .
\eeq
Finally for the tensor box $ij$ matrix element,
$ (- {1 \over 2 \, \nu} \, 
{ 1 \over \Box^{(0)} } \cdot \Box^{(1)} (h,s) \cdot T^{vac} )_{ij} $, one obtains
\beq
- \, {k_i \, k_j \over k^2} \, {1 \over 2 \nu} \, a^2 \, {8 \over 3} \, {\delta G(t) \over G_0} \, \rho(t) \, {\Gamma \over \omega} \, s + {\mathcal{O}} (k^2) \; .
\eeq
The above expressions are now inserted in the general effective field equations of
Eq.~(\ref{eq:field1}), and will give rise to a set of effective field equations appropriate for
this particular gauge, to first order in the field perturbation and with the effects
of $G(\Box)$ included.

%\newpage

\vskip 40pt
\subsection{Field equations in the comoving, synchronous and conformal Newtonian gauges}
\hspace*{\parindent}
\label{sec:efe}

As a result of the previous manipulations one obtains in the comoving gauge
with fields $ (h, s) $ the following $tt$, $ti$, $ii$ (or $xx+yy+zz$), and $ij$ field
equations
\beq
{k^2 \over 3 \, a^2} \left(h + s \right) 
+ {\dot{a} \over a } \, {\dot{h}} 
= 8 \pi G_0 \left(1 + {\delta G \over G_0} \right) \, {\bar{\rho}} \, \delta 
+ 8 \pi G_0 \, {\delta G \over G_0} \, {c_h \over 2 \nu} \, h \, {\bar{\rho}} 
+ {\mathcal{O}} (k^2)
\label{eq:efe_00_comov}
\eeq
\beq
- {1 \over 3} \left( \dot{h} + \dot{s} \right) 
= 8 \pi G_0 \, {\delta G \over G_0} \left( - {1 \over 2 \nu} \right) 
\, {2 \over 9} \, {1 \over i \omega} \, \left( h - 2 s \right)\, {\bar{\rho}} + {\mathcal{O}} (k^2)
\label{eq:efe_0i_comov}
\eeq
\beq
- \, {1 \over 3} {k^2 \over a^2} \, \left( h + s \right) 
- 3 {\dot{a} \over a} \, \dot{h} - \ddot{h} 
= 24 \pi G_0 \, {\delta G \over G_0} \, w_{vac} \, \bar{\rho} \, \delta 
+ 24 \pi G_0 \, {\delta G \over G_0} \, w_{vac}\,{c_h \over 2 \nu} \, h \bar{\rho} 
+ {\mathcal{O}} (k^2)
\label{eq:efe_ii_comov}
\eeq
\beq
{1 \over 6} \, {k^2 \over a^2} \, \left( h + s \right) 
- {3 \over 2 } \, {\dot{a} \over a} \dot{s} 
- {1 \over 2} \, \ddot{s} 
= - \, 8 \pi G_0 \, {\delta G \over G_0} \, {c_s \over 2 \nu} \, s \, \bar{\rho} 
+ {\mathcal{O}} (k^2) \; .
\label{eq:efe_ij_comov}
\eeq
As in Ref. \cite{ht10}, we have found it convenient to here to set in the above
expressions 
\beq
c_s \; \equiv \; \, \left( {8 \over 3} \right) \, {\Gamma \over \omega}
\label{eq:cs}
\eeq
and
\beq
c_h \; = \; \equiv (-1) \, \left( - {11 \over 3} \right) \, {\Gamma \over \omega} 
\;  = \;  \, {11 \over 3}\, {\Gamma \over \omega}  \; .
\label{eq:ch}
\eeq
In the field equations listed above the terms $ {\mathcal{O}} (k^2) $ arise because 
of terms ${\mathcal{O}} (k^2) $ in the expansion of the tensor box operator.

The next step is to convert the left-hand sides of the above field equations, namely
Eqs.~(\ref{eq:efe_00_comov}), (\ref{eq:efe_0i_comov}), (\ref{eq:efe_ii_comov}) and (\ref{eq:efe_ij_comov}), which are all 
expressed in the comoving gauge $ (h, s) $, to the synchronous gauge
with fields $ (h_{sync}, \eta)$.
The result of this change of gauge is the sequential replacement
\bea
{k^2 \over 3 \, a^2} \left(h + s \right) 
+ {\dot{a} \over a } \, {\dot{h}} 
& \;\longrightarrow\; &
- 2 \, {k^2 \over a^2 } \eta + {1 \over a^2} \, {\dot{a} \over a} \, \dot{h}_{sync} 
\nonumber \\
- {1 \over 3} \left( \dot{h} + \dot{s} \right) 
& \;\longrightarrow\; &
2 \, \dot{\eta} 
\nonumber \\
- \, {1 \over 3} {k^2 \over a^2} \, \left( h + s \right) 
- 3 {\dot{a} \over a} \, \dot{h} - \ddot{h} 
& \;\longrightarrow\; &
2 \, {k^2 \over a^2 } \, \eta 
- {1 \over a^2} \, \ddot{h}_{sync}
- 2 \, {1 \over a^2 }\, {\dot{a} \over a} \, \dot{h}_{sync}
\nonumber \\
{1 \over 6} \, {k^2 \over a^2} \, \left( h + s \right) 
- {3 \over 2 } \, {\dot{a} \over a} \dot{s} 
- {1 \over 2} \, \ddot{s} 
& \;\longrightarrow\; &
- \, {k^2 \over a^2 } \, \eta 
+ {1 \over 2} \, {1 \over a^2} \, \left( \ddot{h}_{sync} + 6 \, \ddot{\eta} \right) 
+ {1 \over a^2} \, {\dot{a} \over a} \, \left( \dot{h}_{sync} + 6 \, \dot{\eta} \right) 
\eea
The next step involves one more transformation, this time
from the synchronous $(h_{sync}, \eta)$ to the desired
conformal Newtonian $(\phi,\psi)$  gauge,
\bea
{1 \over a^2} \left[
- 2 \, k^2 \, \eta + {\dot{a} \over a} \, \dot{h}_{sync} 
\right]
& \longrightarrow &
- \, {2 \over a^2} \left[
k^2 \, \phi 
+ 3 \, {\dot{a} \over a} \, \left( \dot{\phi} 
+ { \dot{a} \over a} \, \psi \right) 
\right]
\nonumber \\
2 \, \dot{\eta} 
& \longrightarrow &
2 \, \left( \dot{\phi} + {\dot{a} \over a} \, \psi \right)
\nonumber \\ 
{1 \over a^2 } \left[
2 \, k^2 \, \eta 
- \ddot{h}_{sync}
- 2 \, {\dot{a} \over a} \, \dot{h}_{sync}
\right]
& \longrightarrow &
{6 \over a^2} \left[
\, \ddot{\phi} 
+ {\dot{a} \over a} \, \left( \dot{\psi} + 2 \, \dot{\phi} \right) 
+ \left( 2 \, {\ddot{a} \over a}
- {\dot{a}^2 \over a^2} \right) \, \psi 
+ {k^2 \over 3} \left( \phi - \psi \right) 
\right]
\nonumber \\
{1 \over a^2 } \left[
- \, k^2 \, \eta 
+ {1 \over 2} \, \left( \ddot{h}_{sync} + 6 \, \ddot{\eta} \right) 
+ {\dot{a} \over a} \, \left( \dot{h}_{sync} + 6 \, \dot{\eta} \right)
\right] 
& \longrightarrow &
- \, {k^2 \over a^2} \, \left( \phi - \psi \right) \; .
\eea
Equivalently, the above sequence of two transformations can be described by a single
transformation, from comoving $(h,s)$ to conformal Newtonian $(\phi,\psi)$ gauge,
which is trivially obtained by combining the previous two.
The final outcome of all these manipulations is to achieve a rewrite of
the full set of four original field equations, given in 
Eqs.~(\ref{eq:efe_00_comov}), (\ref{eq:efe_0i_comov}), 
(\ref{eq:efe_ii_comov}) and (\ref{eq:efe_ij_comov}),
now with the left hand side given in the conformal Newtonian gauge 
and the right hand side left in the original comoving gauge.
One obtains
\beq
k^2 \, \phi 
+ 3 \, {\dot{a} \over a} \, \left( \dot{\phi} 
+ { \dot{a} \over a} \, \psi \right) 
= - \, 4 \pi G_0 \, a^2 \, \left( 1 + {\delta G \over G_0} \right) \, \bar{\rho} \, \delta 
- 4 \pi G_0 \, a^2 \, {\delta G \over G_0} \, {c_h \over 2 \nu} \, h \, \bar{\rho} \;
+ \; {\mathcal{O}} (k^2)
\label{eq:efe_00_cN}
\eeq
\beq
\left( \dot{\phi} + {\dot{a} \over a} \, \psi \right) 
= 4 \pi G_0 \, {\delta G \over G_0} \, \left( - \, {1 \over 2 \nu} \right) 
\, {2 \over 9} \,{1 \over i \omega} \, \left(h - 2 \, s \right) \, \bar{\rho} + \; {\mathcal{O}} (k^2)
\label{eq:efe_0i_cN}
\eeq
\bea
\ddot{\phi} 
+ {\dot{a} \over a} \, \left( \dot{\psi} + 2 \, \dot{\phi} \right) 
+ \left( 2 \, {\ddot{a} \over a} 
- {\dot{a}^2 \over a^2} \right) \, \psi 
+ { k^2 \over 3 } \left( \phi - \psi \right) 
& = & 4 \pi G_0 \, a^2 \, \left( w + w_{vac} \, {\delta G \over G_0} \right) \, \bar{\rho} \, \delta \nonumber \\ 
& + & 4 \pi G_0 \, a^2 \, {\delta G \over G_0} \, w_{vac} \, {c_h \over 2 \nu} \, h \, \bar{\rho} \nonumber \\
& + & \; {\mathcal{O}} (k^2)
\label{eq:efe_ii_cN}
\eea
\beq
k^2 \, \left( \phi - \psi \right) 
= + \, 8 \pi G_0 \, a^2 \, {\delta G \over G_0} \, {c_s \over 2 \nu} \, s \, \bar{\rho} \;
+ \; {\mathcal{O}} (k^2) \; .
\label{eq:efe_ij_cN}
\eeq
Note that we have, for convenience, multiplied out the first, third and fourth equations
by a factor of $a^2$.
The last equation involves the quantity 
\beq
\sigma \; = \; { 2 \over 3 } \, {\delta G \over G_0} \, {c_s \over 2 \nu} \cdot  s \; .
\label{eq:sigma}
\eeq
For the purpose of computing the gravitational slip function $\eta \equiv \psi / \phi - 1$
it will be useful here to record the following relationship between perturbations in the
comoving and conformal Newtonian gauge. 
One has
\beq
\psi = - \, { 1 \over 2 k^2 } \, a^2 
\, \left( \ddot{s} + 2 \, {\dot{a} \over a} \, \dot{s} \right)
\label{eq:psi_transf_cN_pbls}
\eeq
\beq
\phi = - \, {1 \over 6} \, \left( h + s \right) 
+ {1 \over 2} \, {a^2 \over k^2} \, {\dot{a} \over a} \, \dot{s}
\label{eq:phi_transf_cN_pbls}
\eeq
Use has been made here of the following relationship between derivatives of
an arbitrary function $f$ in the synchronous and comoving gauges
\beq
{\dot{f}}^{sync} = a {\dot{f}}^{com}
\label{eq:dot_sync_pbls_1}
\eeq
and 
\beq
{d \over d \tau_{sync}} = a \, {d \over d \tau_{com}}
\label{eq:dot_sync_pbls_2}
\eeq
so that
\beq
{\ddot{f}}^{sync} = a^2 \, \left( {{\dot{a}}^{com} \over a} \, {\dot{f}}^{com} + {\ddot{f}}^{com} \right) \; .
\label{eq:dotdot_sync_pbls}
\eeq

%\newpage

\vskip 40pt
\newsection{Gravitational slip function}
\hspace*{\parindent}
\label{sec:slipfcn}

The gravitational slip function is commonly defined as 
\beq
\eta \; \equiv \; { \psi - \phi \over \phi } \; .
\label{eq:eta_def}
\eeq
In classical GR one has $\phi = \psi $ so that $\eta =0$, which makes the quantity $\eta$ a
useful parametrization for deviations from classical GR, whatever their origin might be.
Using the $ij$ field equation given in Eqs.~(\ref{eq:efe_00_cN}), (\ref{eq:efe_0i_cN}),
(\ref{eq:efe_ii_cN}) and (\ref{eq:efe_ij_cN}),  and the relationship between the conformal 
Newtonian fluctuation $ \phi $ and the comoving gauge fluctuations $ h $ and $ s $, 
one finally obtains the rather simple result 
\beq
\eta \equiv { \psi - \phi \over \phi } 
\; = \; - \, 16 \pi G_0 \, {\delta G \over G_0} \, {c_s \over 2 \nu} \, 
{a \over \dot{a}} \, {s \over \dot{s}} \, \bar{\rho} \; .
\label{eq:eta_cs}
\eeq
The last expression contains the quantity
\beq
c_s \; = \; \, \left( {8 \over 3} \right) \, {i \Gamma \over i \omega_s}
\eeq
where $\omega_s$ is the frequency associated with the $s$ perturbation, and
we have made use of $ i \Gamma \rightarrow \dot{a} / a $.
An equivalent form for the expression in Eq.~(\ref{eq:eta_cs}) is 
\beq
\eta \; = \; 
 - \, 16 \pi \, G_0 \, {\delta G \over G_0} \, {1 \over 2 \nu} \, {8 \over 3} 
\, {1 \over i \omega_s} \, {s \over \dot{s}} \, \bar{\rho} 
\; = \; - \, 16 \pi \, G_0 \, {\delta G \over G_0} \, {1 \over 2 \nu} \, {8 \over 3} 
\, { \int \! s \, \mathrm{d}t \over \dot{s}} \, \bar{\rho} \; .
\label{eq:eta_t}
\eeq
In the last expression we now can make use of the equation of motion for the
perturbation $s(t)$ to the order we are working, namely
\beq
\ddot {s}  \, + \, 3 \, { \dot{a} \over a } \, \dot{s} \, = \, 0 \; .
\label{eq:s_t}
\eeq
Let us look here first at the very simple limit of $\lambda \simeq 0$; 
the physically more relevant case of nonzero $\lambda$ will be discussed a
bit later.
Note that, in view of Eq.~(\ref{eq:xi_lambda}),  this last limit 
corresponds therefore to a very large $\xi$.
Then for a perfect fluid with equation of state $p= w \rho $ one has
simply  $a(t) = a_0 (t/t_0 )^{2/3(1+w)} $ and  $\rho (t) = 1 / [6 \pi G t^2 (1+w)^2 ] $, and
from Eqs.~(\ref{eq:eta_t}) or (\ref{eq:eta_a}) one obtains for $w=0$
\beq
\eta \; = \; 4 \cdot { 8 \over 3 } \, c_t \, \left ( { t \over \xi } \right )^3 \, \ln \left ( { t \over \xi } \right )
+ {\cal O} (t^4) 
\eeq
whereas for $w \neq 0 $ one has
\beq
\eta \; = \; 2 \cdot { 8 \over 3 } \, { c_t \over w \, (1-w) } \, \left ( { t \over \xi } \right )^3 \, 
+ {\cal O} (t^6)  \; .
\eeq
Another extreme, but nevertheless equally simple, case is a pure cosmological constant 
term (no matter of any type), which can be modeled by the choice $w=-1$.
In this case $t$ is related to the scale factor by
\beq
{ a(t) \over a_0 } \; = \; \exp \left  \{  \sqrt{\lambda \over 3 } \, (t - t_0 ) \right \} \; .
\eeq
Then, using the relation in Eq.~(\ref{eq:xi_lambda}), one obtains
\beq
{ t \over \xi } \; = \; 1 + \ln { a \over a_\xi } \; ,
\eeq
where the quantity $a_\xi$ is therefore related to the time $t_0$ ("today", $a_0=1$) 
and the scale $\xi$ by
\beq
{  t_0 \over \xi } \; = \; 1 + \ln { 1 \over a_\xi } \; .
\eeq
Since numerically $t_0$ is close to, but smaller than, $\xi$, the scale factor
$a_\xi$ will be close to, but slightly larger than, one.

To actually come up with a definite number for $\eta$ in more realistic cases,
one needs (apart from including
the effects of $\lambda \neq 0 $, which is done below) a value for the coefficient $c_t$
appearing in Eq.~(\ref{eq:grun_t}) for $G(t)$, which in turn is related to the original
expression for the running Newton's constant $G(\Box)$ in Eq.~(\ref{eq:grun_box}).
This issue will be discussed in some detail later, but here let us say the following.
In Ref. \cite{hw05} it was estimated that the values of $c_t$ in Eq.~(\ref{eq:grun_t})
and $c_0$ in Eq.~(\ref{eq:grun_t}) are of the same order of magnitude, $ c_t \approx 0.62 c_0 $.
The most difficult part has been therefore a reliable estimate of $c_0$, which is
obtained from a lattice computation of invariant curvature correlations at fixed geodesic
distance \cite{cor94}, and which, after reexamination of various systematic uncertainties,
leads to the recent estimate used in\cite{ht10} of $c_0 \approx 33.3 $.
That would give $c_t \approx 20.6 $ which, as we will see later, is still very large.
Nevertheless it is expected that $c_0$ (or $c_t$) enter 
{\it all} calculations with $G(\Box)$  with the {\it same} magnitude and sign.

Let us now go back to the more physical case of $\lambda \neq 0$.
The relevant expression for $\eta(t) $ is Eq.~(\ref{eq:eta_t}), where
we use the equation for $ s(t) $, Eq.~(\ref{eq:s_t}), to eliminate the latter.
It is also convenient at this stage to change variables from $t$ to $a(t)$, and use 
the equivalent equation for $s(a)$, namely
\beq
s'' (a) \, + \, \left ( { H' (a) \over H(a) } + { 4 \over a } \right ) \,  s' (a) \; = \; 0
\label{eq:s_a} \; ,
\eeq
where the prime denotes differentiation with respect to the scale factor $a$.
In the above equation one can use, for nonrelativistic matter with equation 
of state such that $w=0$, and to the order needed here, the first Friedmann equation
\beq
H(a) \; = \; \sqrt{ { \lambda \over 3 } + { 4 \over 9 \, a^3 } } \; .
\label{eq:H_a}
\eeq
We have also made use of the unperturbed result for the background matter 
density valid for $ w = 0 $ (which follows from energy conservation), namely
\beq
\bar{\rho} = \bar{\rho}_0 \, {1 \over a^3} \; .
\label{eq:rhobar}
\eeq
Note that the above expression for $ \bar{\rho} $ is valid to zeroth order in $ \delta G $, 
which is entirely adequate when substituted into $ \eta (a) $,  since the rest there is already 
first order in $ \delta G $.
This finally gives an explicit solution for $ s(a) $
\beq
s(a) \; \propto \; { 2 \over 3 a^{3/2} } \, \sqrt { 1 + a^3 \, \theta } \; ,
\eeq\
with parameter $\theta \equiv \lambda / 8 \pi G_0 \, \bar{\rho}_0 $.
The above solution for $s(a)$ can then be substituted directly in Eq.~(\ref{eq:eta_t}),
provided one changes variables from $t$ to $a(t)$, and in the process uses
the following identities
\beq
\int \! s(t) \, \mathrm{d}t = \int \! s(a) \, {1 \over a \, H (a) } \, \mathrm{d}a  \; ,
\eeq
as well as
\beq
\dot{s} = a \, H (a) \, {\partial s \over \partial a} \; ,
\eeq
with $H(a)$ given a few lines above.

The resulting expression, which still involves an integral over the scale factor $a(t)$, can now 
be readily evaluated, and leads eventually to a rather simple expression for $\eta$.
The general result for nonrelativistic matter ($w=0 $) but $\lambda \neq 0 $ is
\beq
\eta (a) \; = \; {16 \over 3 \, \nu} \, {\delta G (a) \over G_0} \, \log \! \left[{a \over a_\xi}\right]  \; .
\label{eq:eta_a}
\eeq
This is the main result of the paper.
The integration constant $a_\xi$ has been fixed following the requirement that
the scale factor $a \rightarrow a_\xi$ for $t \rightarrow \xi $ [see Eqs.~(\ref{eq:grun_box}),
(\ref{eq:grun_t}) and (\ref{eq:grun_a}) for the definitions of $\xi$].
In other words, by switching to the variable $a(t)$ instead of $t$, the quantity $\xi$ has
been traded for $a_\xi$. 
In the next section we will show that in practice the quantity $a_\xi$ is generally
expected to be slightly larger than the scale factor "today", i.e. for $t=t_0$.
As a result the correction in Eq.~(\ref{eq:eta_a}) is expected to be negative today.

The next section will be devoted to establishing the general relationship between
$t$ and $a(t)$, for nonvanishing cosmological constant $\lambda$, so that a 
quantitative estimate for the slip function $\eta$
can be obtained from Eq.~(\ref{eq:eta_a}) in a realistic cosmological context.
Specifically we will be interested in the value of $\eta$ for a current matter fraction 
$\Omega \simeq 0.25$, as suggested by current astrophysical measurements.

\vskip 40pt
\subsection{Relating the scale factor $a$ to $t$, and vice versa}
\hspace*{\parindent}
\label{sec:a0t}

Let us now come back to the general problem of estimating  $\eta (a)$, using the
expression given
in Eq.~(\ref{eq:eta_a}),  for $\lambda \neq 0 $ and a nonrelativistic fluid with
$w=0$.
To predict the correct value for the slip function $\eta (a) $ one needs the 
quantity $\delta G (a)$, which is obtained from the FLRW version of $G(\Box)$, 
namely $G(t)$ in Eq.~(\ref{eq:grun_t}), via the replacement, in this
last quantity, of $ t \rightarrow t(a) $.
The last step requires therefore that the correct relationship between $t$ and $a(t)$
be established, for any value of $\lambda$.
In the following we will first relate $t$ to $a(t)$, and vice versa, to zeroth order in 
the quantum correction $\delta G$ [we will call them $a^{(0)} (t) $ and $t^{(0)} (a) $],
and then compute the first order correction in $\delta G$ to the above quantities
[we will call those $a^{(1)} (t) $ and $t^{(1)} (a) $].

Let us look first at the zeroth order result.
The field equations and the energy conservation equation for $ a^{(0)} (t) $,
without a $ \delta G $ correction, but with the $\lambda $ term, were already given
in Eq.~(\ref{eq:fried_0}),
\bea
3 \, {{\dot{a}}^{(0) \, 2} (t) \over {a}^{(0) \, 2} (t)} 
& = & 
8 \pi \, G_0 \, \bar{\rho}^{(0)} (t) + \lambda 
\nonumber \\
{{\dot{a}}^{(0) \, 2} (t) \over {a^{(0)}}^2 (t)} + 2\, {\ddot{a}^{(0)} (t) \over a^{(0)} (t)} 
& = &
- 8 \pi \, G_0 \, w \, \bar{\rho}^{(0)} (t) + \lambda 
\label{eq:efe_a0t}
\eea
for a spatially flat universe $ ( k = 0 )$, and
\beq
{\dot{\bar{\rho}}}^{(0)}(t) + 3 \, ( 1 + w ) \, 
{ {\dot{a}}^{(0)} (t) \over a^{(0)} (t) } \, \bar{\rho}^{(0)} (t) = 0 \; .
\label{eq:encons_a0t}
\eeq
From these one can obtain $ a^{(0)}(t) $ and then $ \bar{\rho}^{(0)}(t) $.
As a result the scale factor is found to be related to time by
\beq
t^{(0)}(a) 
= { 2 \, \mathrm{Arcsinh} \! \left[ \, a^{3/2} \, \theta^{1 \over 2} \right] 
\over \sqrt{3 \, \lambda} }
\label{eq:t0_a}
\eeq
where we have defined the parameter
\beq
\theta \; \equiv \; { \lambda \over 8 \pi G_0 \bar \rho_0 } \; = \; { 1 - \Omega \over \Omega }
\label{eq:theta}
\eeq
with $\bar \rho_0$ the current ($t=t_0$) matter density, and $\Omega$ the
current matter fraction.
Note that in practice we will be interested in a matter fraction which
today is around $0.25$, giving $\theta \simeq 3.0 $, a number which
is of course quite far from the zero cosmological constant case of $\theta =0$.

One can express the time today ($t_0$) in terms of cosmological constant $ \lambda $,
and therefore in terms of $ \theta $, as follows
\beq
t_0^{(0)} = { 2 \, \mathrm{Arcsinh}({\sqrt{\theta}}) \over \sqrt{3 \lambda} }
\label{eq:t0}
\eeq
with the normalization for $ t^{(0)}(a) $ such that 
$ t^{(0)}(a=0) = 0 $ and $ t^{(0)}(a=1) = t_0 $ "today".
So here we follow the customary choice of having the scale factor equal to one "today".
Then one has 
\beq
{t^{(0)}(a) \over t_0^{(0)}} 
={\mathrm{Arcsinh} \! \left[ \sqrt{a^{3} \, \theta} \right] 
\over 
\mathrm{Arcsinh}({\sqrt{\theta}})} \; .
\label{eq:t0_a_norm}
\eeq
When expanded out in $\theta$, the above result leads to some perhaps
more recognizable terms,
\beq
{t^{(0)}(a) \over t_0^{(0)}} 
= a^{3 \over 2} 
\, \left[ 1 - {1 \over 6}\, \left(-1 + a^3 \right) \, \theta + {1 \over 360}\, \left(-17 - 10 \, a^3 + 27 \, a^6 \right) \, \theta^2 
+ \cdots \right] \; .
\label{eq:t0_a_power}
\eeq
Conversely, one has for the scale factor as a function of the time
\beq
a^{(0)}(t) = \left({ \mathrm{Sinh}^2 \! \left[{\sqrt{3 \, \lambda} \over 2} \, t\right] \over \theta }\right)^{1 \over 3} \; ,
\label{eq:a0t_norm}
\eeq
which, when expanded out in $\lambda$ or $t$, gives the more recognizable result
\beq
\left [ a^{(0)}(t) \right ]^3 \; = \; 
{3 \, \lambda \, t^2 \over 4 \theta} \, 
\left(1 + {\lambda \, t^2 \over 4} + {\lambda^2 \, t^4 \over 40} + \cdots \right) \; .
\eeq
Similarly for the pressure one obtains
\beq
\bar{\rho}^{(0)}(t) 
= { \lambda \, \mathrm{Csch}^2 \! \left[ {\sqrt{3 \, \lambda} \over 2} \, t \right] \over 8 \pi G_0} \; ,
\label{eq:rho0t}
\eeq
which when expanded out in $\lambda$ or $t$ gives the more familiar result
\beq
\bar{\rho}^{(0)}(t) 
= {1 \over 6 \pi G_0 \, t^2 
\, \left(1 + {t^2 \, \lambda \over 4} + {t^4 \, \lambda^2 \over 40} + {3 \, t^6 \, \lambda^3 \over 2240} + \cdots \right)} \; .
\eeq
To be more specific, let us set $\theta=3$, which corresponds to a matter fraction today
of $\Omega \sim 0.25$.
In addition, we will now make use of Eq.~(\ref{eq:xi_lambda}) and set 
$\lambda \rightarrow {3/\xi^2} $.
One then obtains
\beq
t_0^{(0)} \left(\theta = 3 \right) = 0.878 \;  \xi \; ,
\label{eq:t0_theta3}
\eeq
which shows that $t_0$ and $\xi$ are rather close to each other
(apparently a numerical coincidence).

Then, from the expression for $G(t)$ in Eq.~(\ref{eq:grun_t}),
\beq
{\delta G (t) \over G_0} = c_t \left( {t \over \xi} \right)^{1 \over \nu} \; ,
\eeq
one can obtain $G(a)$ in all generality, by the replacement $t \rightarrow t(a)$
according to the result of Eqs.~(\ref{eq:t0_a}) or (\ref{eq:t0_a_norm}).
For the special case of pure nonrelativistic matter with equation of state $w=0$ and $\lambda =0$
one obtains, using Eq.~(\ref{eq:t0_a_power}),
\beq
{\delta G (a) \over G_0} \; = \;  c_t \left( {a \over a_\xi} \right)^{\gamma_\nu} \; ,
\eeq
with exponent
\beq
\gamma_\nu = {3 \over 2 \nu} \; .
\eeq
The latter is largely the expression used earlier in the matter density perturbation
treatment of our earlier work of Ref. \cite{ht10}.

More generally one can define $ a_\xi $ as the value for the scale factor $a$ which
corresponds to the scale $ \xi $,
\beq
a_\xi^{(0)}  \; \equiv \; 
\left( {1 \over \theta} \right)^{1\over 3} \mathrm{Sinh}^{2 \over 3} \! \left[ {3 \over 2}\right] 
= 1.655 \, \left( {1 \over \theta} \right)^{1\over 3} \; ,
\label{eq:a_xi}
\eeq
so that in general $ a_\xi \neq a_0 $, where $ a_0 =1 $ is the scale factor "today".
Then for the observationally favored case $ \theta \simeq 3 $ one obtains
\beq
a_\xi^{(0)}(\theta = 3) = 1.148 \; ,
\label{eq:axi_a0}
\eeq
which clearly  implies $ a_\xi^{(0)}  > a_0 = 1 $.
\footnote{
Let us give here a few more observational numbers for present and future reference. 
From the present age of the Universe
$t_0 \approx 13.75 \, Gyrs \simeq 4216 \, Mpc $, whereas from the observed 
value of $\lambda $ (mostly extracted from distant supernovae surveys)
one has following Eq.~(\ref{eq:xi_lambda}) $ \xi \simeq  4890 \, Mpc $, 
which then gives $t_0 / \xi \simeq 0.862  = 1 / 1.160 $.
This last ratio is similar to the number we used in Eq.~(\ref{eq:t0_theta3}),
by setting there $\Omega=0.25$ exactly. 
}
The above expressions will be used in the next section to obtain a quantitative
estimate for the slip function $\eta (a)$, evaluated at today's time $t=t_0$.

% \vskip 40pt
% \subsection{$ a^{(1)}(t) $ and $ \rho^{(1)}(a) $}
% \hspace*{\parindent}
% \label{sec:a1t}

The discussion above dealt with the case of $\delta G=0$.
Let us now consider briefly the corrections to $a(t)$ and, conversely, $t(a)$
that come about when the running of $G$ is included, in other words when 
a constant  $G$ is replaced by $G(t)$ or $G(a)$ in the effective field equations.
In Eq.~(\ref{eq:fried_run}) the Friedman equations were given in the presence
of a running $G$, namely
\bea
3 \, {{\dot{a}}^2 (t) \over {a}^2 (t)} 
& = & 
8 \pi \, G_0 \left ( 1 + {\delta G(t) \over G_0} \right ) \, \bar{\rho}(t) + \lambda 
\nonumber \\
{{\dot{a}}^2 (t) \over {a}^2 (t)} + 2\, {\ddot{a} (t) \over a (t)} 
& = & 
- 8 \pi \, G_0 \, \left ( w + w_{vac} {\delta G(t) \over G_0} \right )\, \bar{\rho} (t) + \lambda  \; ,
\label{eq:fried_run1}
\eea
together with the energy conservation equation
\beq
3 \, {\dot{a} (t) \over a (t)} \, 
\left [ \left (1+w \right ) + \left (1+ w_{vac} \right )\,{\delta G(t) \over G_0} 
\right ] \bar{\rho} (t) 
+ { \dot{\delta G}(t) \over G_0} \, \bar{\rho} (t)
+ \left ( 1 + {\delta G(t) \over G_0} \right )\, \dot{\bar{\rho}} (t) = 0 \; .
\label{eq:encons_zeroth_w_1}
\eeq
To solve these equations to first order in $\delta G$ we set
\beq
a(t) = a^{(0)} (t) \left [ 1 + c_t \, a^{(1)}(t) \right ]
\label{eq:a0t_1t}
\eeq
\beq
\bar{\rho}(t) = \bar{\rho}^{(0)} (t) \left [ 1 + c_t \,\bar{\rho}^{(1)} (t)\right ]
\label{eq:rho0t_1t}
\eeq
where $ a^{(0)} (t) $ and $ \bar{\rho}^{(0)} (t) $ here represent the solutions
obtained previously for $\delta G=0$.
One then finds for the correction to the matter density
\beq
\bar{\rho}^{(1)} (t) = - \, \left( { t \over \xi }\right)^{1 \over \nu} \, 
\left(1 + w_{vac} \, {\nu \over (1 + \nu)} 
\, \sqrt{3 \, \lambda} \; t \;\; \mathrm{Coth} \! \left[{ \sqrt{3 \, \lambda} \over 2} \, t \right] 
\right)
\label{eq:rho1t}
\eeq
and to lowest nontrivial order in $t$ and for $ w_{vac}=1/3 $ 
\beq
\bar{\rho}^{(1)} (t) = - \, {3 + 5 \nu \over 3 ( 1 + \nu)} \, \left({t \over \xi} \right)^{1 \over \nu} + \dots \; .
\eeq
For the correction to the scale factor one finds
\beq
a^{(1)} (t) = - \, w_{vac} \, {\nu \over (1 + \nu)} \, \lambda \; 
\int^t_0 \! 
{t^{\prime} \left({ t^{\prime} \over \xi }\right)^{1 \over \nu} 
\over 
- 1 + \mathrm{Cosh} \! \left[ \sqrt{3 \, \lambda} \; t^{\prime} \right]} \, 
\mathrm{d}t^{\prime}
\label{eq:a1t}
\eeq
and to lowest nontrivial order in $t$ for $ w_{vac}=1/3 $,
\beq
a^{(1)} (t) = - \, {2 \nu^2 \over 9 \, ( 1 + \nu)} \, \left({t \over \xi} \right)^{1 \over \nu} + \dots \; .
\eeq
After having obtained the relevant formulas for $a(t)$ and $t(a)$ in the general case, i.e. for
nonzero $\lambda$, we can return to the problem of evaluating the slip function $\eta$.

\vskip 40pt
\subsection{Quantitative estimate of the slip function $\eta (z) $}
\hspace*{\parindent}
\label{sec:eta_z}

The general expression for the gravitational slip function 
$\eta (a)$ was given earlier in Eq.~(\ref{eq:eta_a}) 
for $w=0$ and $\lambda \neq 0$,
\beq
\eta (a) \; = \; {16 \over 3 \nu} \, {\delta G (a) \over G_0} \, \log \! \left[{a \over a_\xi}\right]  \; .
\eeq
To obtain $\delta G(a)$ we now use, from Eq.~(\ref{eq:grun_t}), 
\beq
{\delta G (t) \over G_0} = c_t \, \left({t \over \xi} \right)^{1 \over \nu}
\label{eq:grun_t_1}
\eeq
and substitute in the above expression for $\delta G(t) $ the correct relationship between 
$t$ and $a$, namely $t(a)$ from Eq.~(\ref{eq:t0_a}), which among other things
contains the constant defined in Eq.~(\ref{eq:a_xi}),
\beq
a_\xi \; = \; 
\left( {1 \over \theta} \right)^{1\over 3} \mathrm{Sinh}^{2 \over 3} \! \left[ {3 \over 2}\right]  \; .
\label{eq:axi_1}
\eeq
It will be convenient, at this stage, to also make use of the relationship
in Eq.~(\ref{eq:xi_lambda}), namely
\beq
\lambda \; \rightarrow \; {3 \over \xi^2} \; .
\eeq
The last step left is to make contact with observationally accessible quantities,
by expanding  in the redshift $ z $, 
related in the usual way to the scale factor $a$ by $ a \equiv {1 /(1 + z)} $.
Then for $ \nu = 1/3 $ and $ \theta = 3 $ (matter fraction $\Omega=0.25$) one 
finally obtains for the gravitational slip function
\beq
\eta(z) \; = \; - \, 1.491 \,c_t - 6.418 \, c_t \, z + 30.074 \, c_t \, z^2 + \cdots
\label{eq:eta_z}
\eeq
To obtain an actual number for $\eta (z=0) $ one needs to address two more issues.
They are (i) to provide a bound on the theoretical uncertainties in the above expression,
and (ii) to give an estimate for the coefficient $c_t$, which is traced back to Eq.~(\ref{eq:grun_t})
and therefore to the original expression for $G(\Box)$ in Eq.~(\ref{eq:grun_box}).
The latter contains the coefficient $c_0$, but in Ref. \cite{hw05} the estimate
was given $c_t = 0.450 \; c_0$ for the tensor box operator; thus $c_t$ and $c_0$
can safely be assumed to have the same sign, and comparable magnitudes.

To estimate the level of uncertainty in the magnitude of the correction coefficient in
Eq.~(\ref{eq:eta_z}) we will consider here an infrared regulated version
of $G(\Box)$, where an infrared cutoff is supplied so that in Fourier space $k > \xi^{-1}$,
and the spurious infrared divergence at small $k$ is removed.
This is quite analogous to an infrared regularization used very successfully 
in phenomenological applications to QCD heavy quark bound states \cite{ric79,eic81}, and 
which has recently found some limited justification in the framework of infrared 
renormalons \cite{ben99}.
As shown already in the first cited reference, it works much better than expected;
here a similar prescription will be used just as a means to provide some estimate
on the theoretical uncertainty in the result of Eq.~(\ref{eq:eta_z}).
Therefore, instead of the $G(\Box)$ in Eq.~(\ref{eq:grun_box}),
which in momentum space corresponds to 
\begin{equation}
G(k^2) \; \simeq \; G_0 \, \left [ 
\, 1 \, + \, c_0 \, \left ( { 1 \over \xi^2 \, k^2 } \right )^{ 1/ 2 \nu }
\, + \, \dots \right ] \; ,
\label{eq:grun_k} 
\end{equation}
we will consider a corresponding infrared regulated version,
\beq
G(k^2) \; \simeq \; G_0 \left [ \; 1 \, 
+ \, c_0 \left ( { \xi^{-2} \over k^2 \, + \, \xi^{-2} } \right )^{1 / 2 \nu} \, 
+ \, \dots \; \right ] \; .
\label{eq:grun_k_reg}
\eeq
Of course the small distance, $  k \gg \xi^{-1} $ or $ r \ll \xi $, behavior is unchanged,
whereas for large distances $r \gg \xi$  the gravitational coupling
no longer exhibits the spurious infrared divergence;
instead  it approaches a finite value $G_\infty \simeq ( 1 + c_0 + \dots ) \, G_0 $.
Now, in momentum space the infrared regulated $ \delta G (k) $ reads
\beq
{\delta G(k^2) \over G_0} = c_0 \, \left({m^2 \over k^2 + m^2} \right)^{1 / 2 \, \nu} \; ,
\label{eq:grun_k_reg_1}
\eeq
with $m= 1/ \xi $, and in position space the corresponding form is
\beq
{\delta G(\Box) \over G_0} = c_0 \, \left({1 \over - \, \xi^2 \Box + 1 } \right)^{1 / 2 \, \nu} \; .
\label{eq:grun_box_reg}
\eeq
Following the results of Ref. \cite{hw05}, if the above differential operator 
acts on functions of $ t $ only, then one obtains for $\delta G (t) $
\beq
{\delta G (t) \over G_0} \; = \;  c_0 \, \left ( 
{1 \over \left({c_0 \over c_t}\right)^{2 \, \nu} \, \left({\xi \over t}\right)^2 + 1} 
\right)^{1 \over 2 \, \nu}
\label{eq:grun_t_reg}
\eeq
with again $ c_t / c_0 \approx 0.62 $ \cite{hw05}.
Note that the expression in Eq.~(\ref{eq:grun_t_reg}) could also have been obtained
directly from Eq.~(\ref{eq:grun_t}), by a direct regularization.

One can then repeat the whole calculation for $\eta (a)$ with the regulated
version of $\delta G(t)$ given in Eq.~(\ref{eq:grun_t_reg}).
The result is 
\beq
\eta (z) = - 0.766 \, c_t - 4.109 \, c_t \, z + 12.188 \, c_t \, z^2 + \cdots \; .
\label{eq:eta_z_reg}
\eeq
It seems that the effect of the infrared regularization has been to
reduce the magnitude of the effect (at $z=0$) by about a factor of 2.
It is encouraging that, at this stage of the calculation,
the negative trend in $ \eta (z) $ due to the running of 
$ G $ appears unchanged.
Furthermore, in all cases we have looked so far, the value $\eta (z=0)$ is found to
be negative.

\vskip 40pt
\subsection{Slip function $\eta(z)$ for stress perturbation $s = 0$}
\hspace*{\parindent}
\label{sec:eta_s_0}

In Ref. \cite{ht10} a preliminary estimate of the magnitude of the slip
function $\eta$ was given.
The calculation there neglected the stress field $s$ in Eq.~(\ref{eq:stress_def})
and only included the
metric perturbation $h$ in the comoving gauge.
The main reason was that nonrelativistic matter density perturbations,
and therefore the growth exponents, are unaffected by the stress field contribution.
We will show here that in this case one still obtains a nonvanishing $ \eta $, 
whose value we will discuss below.
The results will be useful, since now a direct comparison can be done
with the full answer (including the stress field) for $\eta (z) $ given in the previous section.

In the absence of stress ($s=0$) and finite $k$, the $tt$ and $xx + yy + zz $ field equations
read
\beq
- 2 \, {k^2 \over a^2} \phi 
- 8 \pi G_0 {c_h \over 2 \nu} \, {\delta G \over G_0} \, \rho \; \delta \, \left( - {2 \over 1 + w} \right)
= 8 \pi G_0 \, \left(1 + {\delta G \over G_0} \right) \, \rho \, \delta
\eeq
\beq
2 {k^2 \over a^2} \, \left(\psi - \phi \right) 
+ 24 \pi G_0 {c_h \over 2 \nu} \, w_{vac} \,{\delta G \over G_0} \, \rho \; \delta \, \left( - {2 \over 1 + w } \right)
= - \, 24 \pi G_0 \, \left(w + w_{vac} \, {\delta G \over G_0} \right) \, \rho \, \delta  \; .
\eeq
In both equations we have made use of zeroth order (in ${\delta G/ G_0}$) energy conservation,
which leads to $ h = - {2 \over ( 1 + w )} \, \delta $, where $\delta$ is the matter fraction.
One can then take the ratio of the two equations given above, and obtain 
again an expression for the slip function $ \eta = { \left(\psi - \phi\right)/ \phi } $. 
For $ w =0 $ (nonrelativistic matter), after expanding in $ {\delta G/G_0} $, one finds the
rather simple result
\beq
\eta \; = \; {\psi - \phi \over \phi}  \; = \; 
3 \, w_{vac}  \, \left( 1 - {c_h \over \nu} \right) \, {\delta G \over G_0} \; .
\eeq
Here the quantity $c_h$ is the same as in Eq.~(\ref{eq:ch}), and depends on the
choices detailed below.
In the following we will continue to use $w_{vac} = 1/3 $ [see Eqs.~(\ref{eq:wvac_def}) 
and (\ref{eq:wvac})] \cite{hw05,ht10}, which is the correct value associated with
 $G (\Box)$ in the FLRW background metric.

In Ref. \cite{ht10} we used the {\it scalar box} value $c_h = 1/2$, which then gives
\beq
\eta \; = \; \left( 1 - {1 \over 2 \, \nu} \right) \, {\delta G \over G_0} \; = \; 
\left( 1 - {1 \over 2 \, \nu} \right) \, c_t \, \left({t \over \xi}\right)^{1 \over \nu} 
+ \cdots
\eeq
In this last case it is then easy to recompute the slip function in terms of the redshift, 
just as was done in the previous section, and one finds, 
under the same conditions as before [$ \nu = 1/3 $, $\theta=3$,
and $t_0/ \xi$ as given in Eq.~(\ref{eq:t0_theta3})] the following result
\beq
\eta \simeq - \, 0.338 \, c_t \; + \; O( z) \; .
\eeq
For the infrared regulated version of ${\delta G/G_0}$ given in Eq.~(\ref{eq:grun_t_reg})
one obtains instead the slightly smaller value
\beq
\eta \simeq - \, 0.174 \, c_t \; + \; O( z) \; .
\eeq
For the {\it tensor box} case (also discussed extensively in \cite{ht10}, where it
was shown that this is in fact the correct way of doing the calculation) one finds a
significantly larger value $ c_h \simeq 7.927 $, so that in this case the slip 
function $ \eta $ becomes
\beq
\eta \simeq \left( 1 - {7.927 \over \nu} \right) \, { \delta G \over G_0 } 
= \left( 1 - {7.927 \over \nu} \right) \, c_t \, \left({t \over \xi}\right)^{1 \over \nu} 
+ \cdots
\label{eq:eta_h}
\eeq
Also in this case one can recompute the slip function in terms of the redshift, 
and one finds, under the same conditions as before,
\beq
\eta \simeq - \, 15.42 \, c_t \; + \; O(z) \; .
\label{eq:eta_h_reg}
\eeq
For the infrared regulated $ {\delta G/G_0} $ given in Eq.~(\ref{eq:grun_t_reg})
one finds instead 
\beq
\eta \simeq - \, 7.919 \, c_t \; + \; O(z) \; ,
\eeq
which is again about a factor of 2 smaller than the unregulated value.

We conclude from the above exercise of calculating $\eta$ with vanishing
stress field $s=0$ three things.
The first is that using the scalar box result on the trace of the 
energy-momentum tensor (which ultimately is not an entirely correct, or at least 
an incomplete,  procedure, given the tensor nature of the matter 
energy-momentum tensor) underestimates the effects of $G(\Box)$ on 
the slip function $\eta (z=0)$ by a factor that can be as large as an
order of magnitude.

The second lesson is that the stress field ($s$) contribution is indeed important,
since it reduces the size of the quantum correction significantly 
[Eqs.~(\ref{eq:eta_z}) and (\ref{eq:eta_z_reg})], compared to the $s=0$ 
result [Eqs.~(\ref{eq:eta_h}) and (\ref{eq:eta_h_reg})], again by almost
an order of magnitude, which would imply some degree of cancellation
between the $s$ and $h$ contributions.

The third observation is that in all cases we have looked at so far the quantum
correction to the slip function is {\it negative} at $z=0$.

\vskip 40pt
\newsection{Conclusions}
\hspace*{\parindent}
\label{sec:conclusion}

In the previous sections we computed corrections to the gravitational 
slip function $\eta = \psi / \phi -1 $
arising from the renormalization-group motivated running $G(\Box)$, as 
given in Eq.~(\ref{eq:grun_box}).
The relevant result was presented in Eqs.~(\ref{eq:eta_z}) and (\ref{eq:eta_z_reg}), 
the first expression representing the answer for an unregulated $G(\Box)$, 
and the second answer found for an infrared regulated version of the same.
It should be noted that, so far, in the treatment of metric and matter perturbations 
we have considered only the ${\bf k} \rightarrow 0 $ limit [see Eq.~(\ref{eq:fourier})].
Let us focus here for definiteness on the first of the two results 
[Eq.~(\ref{eq:eta_z})], which is
\beq
\eta (z) \; \simeq \;  - 1.491 \, c_t \; + \; O(z)
\label{eq:eta_z_1}
\eeq
at $z \simeq 0$.
We now come to the last issue, namely an estimate for the magnitude of the constant $c_t$.
As already discussed previously in Sec. \ref{sec:eta_z}, to get an actual number for $\eta (z=0) $ 
one needs a number for $c_t$, whose appearance is traced back to Eq.~(\ref{eq:grun_t}),
and therefore to the original expression for $G(\Box)$ in Eq.~(\ref{eq:grun_box}),
with $c_t  \approx 0.450 \times c_0$ for the relevant tensor box operator \cite{hw05} .

The value of the constant $c_0$ has to be extracted from a nonperturbative lattice 
computation of invariant curvature correlations at fixed geodesic distance \cite{cor94};
it relates the physical correlation length $\xi$ to the bare lattice coupling $G$, and is
therefore a genuinely nonperturbative amplitude.
After a reexamination of various systematic uncertainties, these lead to the recent 
estimate used in \cite{ht10} of $c_0 \approx 33.3 $.
That would give for the amplitude $c_t \approx 20.6 $ which still seems rather large.
Nevertheless, based on experience with other field-theoretic models which
also exhibit nontrivial fixed points such as the nonlinear sigma model, 
as well as QCD and non-Abelian gauge theories,
one would expect this amplitude to be of order unity; 
very small or very large numbers would seem rather atypical and un-natural.

As far as astrophysical observations are concerned,
current estimates for $\eta (z=0)$ obtained from CMB measurements
give values around $0.09 \pm 0.7 $ \cite{ame07,dan09},
which would then imply an observational bound $c_t \lsim 0.3 $.

Indeed a similar problem of magnitudes for the theoretical amplitudes
was found in our recent calculation of matter 
density perturbations with $G(\Box)$, where again the corrections 
seemed rather large \cite{ht10} in view of the above quoted value of $c_t$.
Let us briefly summarize those results here.
Specifically, in Ref. \cite{ht10} a value for the density perturbation growth 
index $\gamma$ was obtained in the presence of $G(\Box)$.
The quantity $\gamma$ is in general obtained from the growth index $ f (a) $ \cite{pee93}
\beq
f (a) \equiv { \partial \ln \delta (a) \over \partial \ln a }  \; ,
\label{eq:fa_def}
\eeq
where $\delta (a)$ is the matter density contrast.
One is mainly interested in the neighborhood of the present era, $a (t) \simeq a_0 \simeq 1 $,
which leads to the definition of the growth index parameter $ \gamma $ via
\beq
\gamma \equiv  \left. { \ln f \over \ln \Omega  }  \right \vert_{a=a_0} \; .
\label{eq:gamma_def}
\eeq
The latter has been the subject of increasingly accurate cosmological observations, for
some recent references see \cite{smi09,vik09,rap09}.
\footnote{
For a recent detailed review on the many tests of general relativity on astrophysical scales,
and a much more complete set of references, see for example \cite{uza02,uza09}.}

On the theoretical side, for the tensor box one finds \cite{ht10}, 
for a matter fraction $\Omega = 0.25$,
\beq 
\gamma =  0.556 - 106.4 \, c_t + O(c_t^2 ) \; .
\label{eq:gamma_ten}
\eeq
where the first contribution is the classical GR value from the relativistic
treatment of matter density perturbations \cite{pee93}.
The result presented above is in fact a slight improvement over the answer quoted in our 
earlier work \cite{ht10},
since now the improved relationship between $t$ and $a$ given in Eq.~(\ref{eq:t0_a})
has been used, which reduces the magnitude of the correction proportional
to $c_t$.
Nevertheless, it should be emphasized that the above result has been obtained
in the ${\bf k} \rightarrow 0 $ limit of the perturbation Fourier modes in Eq.~(\ref{eq:fourier}).

Recent observational bounds on x-ray studies of large galactic clusters at distance scales 
of up to about $1.4$ to $8.5 Mpc$ 
(comoving radii of $\sim 8.5 Mpc $ and  viral radii of $\sim 1.4 Mpc$) \cite{vik09} 
favor values for $\gamma= 0.50 \pm 0.08 $, and more recently $\gamma = 0.55+0.13-0.10 $
\cite{rap09}.
This would then constrain the amplitude $c_t$ in Eq.~(\ref{eq:gamma_ten}) 
{\it at that scale} to $c_t \lsim 5 \times 10^{-4}$.
The latter bound from density perturbations
seems a much more stringent bound than the one
coming from the observed slip function.
Indeed with the bound on $c_t$ coming from the observed density perturbation
exponents one
would conclude that, according to Eq.~(\ref{eq:eta_z_1}), the correction to the slip
function at $z \simeq 0 $ must indeed be very small, $ \eta \simeq O(10^{-3}) $, 
which is a few orders of magnitude below the observational limit quoted above, 
$\eta \simeq 0.09 \pm 0.7 $.

It is of course possible that the galactic clusters in question are not large enough 
yet to see the quantum effect of $G(\Box)$, since after all the relevant scale in 
Eq.~(\ref{eq:grun_box}) is related to $\lambda$ and is
supposed therefore to be very large, $\xi \simeq 4890 Mpc $.
\footnote{
One might perhaps think that the running of $G$ envisioned here might lead to small 
observable consequences on much shorter, galactic length scales.
That this is not the case can be seen, for example, from the following argument.
For a typical galaxy one has a size $\sim 30 \, kpc$, giving for the quantum correction
the estimate, from the potential obtained in the
static isotropic metric solution with $G(\Box)$ \cite{hw06} which gives 
$\delta G(r) \sim (r/\xi)^{1/\nu}$, 
$( 30 \, kpc / 4890 \times 10^3 \, kpc )^3 \sim 2.31 \times 10^{-16} $ which is tiny given
the large size of $\xi$.
It is therefore unlikely that such a correction will be detectable at these length scales,
or that it could account for large anomalies in the galactic rotation curves.
The above argument also implies a certain sensitivity of the results to the value of the scale $\xi$;
thus an increase in $\xi$ by a factor of two tends to reduce the effects of $G(\Box)$ by 
roughly $2^3 = 8$, as can be seen already from Eq.~(\ref{eq:grun_box}) with $\nu=1/3$.
More specifically, the amplitude of the quantum correction is proportional, 
in the non-infrared improved case, to the combination $c_0 / \xi^3$.}
But most likely the theoretical uncertainties in the value of $c_t$ have also been underestimated
in \cite{cor94}, and new, high precision lattice calculation will be required to 
significantly reduce the systematic errors.

Nevertheless it seems clear that the non-perturbative coefficient $c_0$ (or $c_t$) 
enters {\it all} calculations involving $G(\Box)$ with the {\it same} magnitude and sign.
This is simply a consequence of $c_0$ being part of the renormalization
group $G(\Box)$ which enters the covariant effective field equations of Eq.~(\ref{eq:field1}).
Consequently, one should be able to relate one set of physical results to another, 
such as the value of the slip function $\eta (z=0)$ in 
Eq.~(\ref{eq:eta_z}) to the corrections to the density perturbation growth 
exponent $\gamma$  computed in \cite{ht10}, and given here in Eq.~(\ref{eq:gamma_ten}).
Then the amplitude $c_t$ can be made to conveniently drop out when computing
the ratio of $G(\Box)$ corrections to two different physical processes.
The resulting predictions are then entirely independent of the theoretical
uncertainty in the amplitude $c_0$, and remain sensitive only to the uncertainties in
the two other quantum parameters $\xi$ and $\nu$, which are expected to be 
significantly smaller.
One then obtains for the ratio of the corrections to the growth exponent $\gamma$
to the slip function $\eta (z=0)$  at $ z \simeq 0 $ 
\beq
{ \delta \, \gamma \over \delta \, \eta } \; \simeq \; 
{ - 106.4 \, c_t \over  - 1.491 \, c_t } \; \simeq \;  + \, 71.4 
\eeq
for the infrared unimproved case.
One conclusion that one can draw from the numerical value of the above ratio is
that it might be significantly harder to see the $G(\Box)$ correction
in the slip function than in the matter density growth exponent, by almost 2
orders of magnitude in relative magnitude.
Hopefully increasingly accurate astrophysical measurements of the latter will
be done in the not too distant future.
Of particular interest would be any trend in the growth exponents as a function of
the maximum galactic cluster size.

 \newpage

\vspace{20pt}

{\bf Acknowledgements}

One of the authors (HWH) wishes to thank Thibault Damour and Gabriele Veneziano
for inspiration and discussions leading to the present work, and
Alexey Vikhlinin for correspondence regarding astrophysical measurements 
of structure growth indices.
HWH wishes to thank Thibault Damour and the
I.H.E.S. in Bures-sur-Yvette for a very warm hospitality. 
The work of HWH was supported in part by the I.H.E.S. and
the University of California.
The work of RT was supported in part by a DoE GAANN student fellowship.

\vspace{20pt}

\newpage

\appendix

\section*{Appendix}

\newsection{Scalar box in the comoving gauge}
\hspace*{\parindent}
\label{sec:scalarbox}

In this section we will give a short sample calculation of the effects of the
covariant d'Alembertian operator 
$\Box \equiv g^{\mu\nu} \, \nabla_\mu \nabla_nu $  acting on a 
coordinate scalar, such as the trace of the energy-momentum tensor.
The calculation presented below will show that the result is unchanged
when the stress contribution $s$ is included in the metric for the comoving gauge.
Specifically here we will be interested in the correction of order $h_{ij}$ that 
arises when the operator in Eq.~(\ref{eq:gbox_hs_expanded})
acts on the scalar $ T_{\lambda}^{\;\; \lambda} = - {\bar \rho} $.
Thus, for example, it will give the correction $O(h,s)$ to 
$\delta \rho_{vac}$, namely the second term in the expression
\beq
\delta \rho_{vac} (t) = { \delta G(\Box^{(0)}) \over G_0 }\; \delta \rho (t)
+ {\delta G (\Box) (h,s) \over G_0} \; \bar{\rho} (t) \; ,
\label{eq:delta_rhovac_h}
\eeq
with the first term being simply given in the FLRW background
by $\delta G(t)/G_0 \cdot \delta \rho (t) $.
Here the $O(h,s)$ correction is given explicitly by the expression
\beq
{\delta G(\Box) (h,s) \over G_0} \, {\bar \rho} \, = \, 
- {1 \over 2 \, \nu} \, { c_0 \over \xi^{{1 / \nu}} } \,
{ 1 \over \Box^{(0)} } \cdot \Box^{(1)} (h,s) \cdot
\left ( { 1 \over \Box^{(0)} } \right )^{ 1/ 2\nu} \cdot
{\bar \rho} \; .
\eeq
Now the covariant d'Alembertian $\Box $ acting on general 
scalar functions $S(x)$ simplifies to 
\beq
\Box \, S(x) \equiv {1 \over \sqrt{g} } \, 
\partial_\mu \, g^{\mu\nu} \sqrt{g} \, \partial_\nu \, S(x) \; .
\eeq
In the absence of $h_{ij}$ fluctuations this gives for the metric in the comoving gauge
\beq
\Box^{(0)} S(x) = 
{1 \over a^2} \nabla ^2 S - 3\, {\dot{a} \over a} \, \dot{S} - \ddot{S} \; .
\eeq
To first order in the field fluctuation $h_{ij}$ of the comoving  gauge one computes
\bea
\Box^{(1)} (h, s) \, S(x)
& = & \dot{S} \, \left[ - \, {1 \over 2} \dot{h} \right]
+ \; \partial_x S \, \left[ {1 \over 6 a^2} \, i \, k_x \, \left(h + 4 \, s \right)\right]
\nonumber \\
& \; & + \; \partial_x^2 S \, \left[ - \, {1 \over 3 a^2} \, \left(h + s \right) +{1 \over a^2} \, {k_x^2 \over k^2} \, s \right]
+ \partial_x \partial_y S \, \left[{ 2 \over a^2 } \, {k_x \, k_y \over k^2} \, s \right]
\eea
where we have set as usual $h(x) = h(t) \, e^{i \, \bf{k} \cdot \bf{x}}$.
But, for a function of time only, one obtains
\beq
\Box^{(1)} (h) \, \rho (t) = - { 1 \over 2} \, \dot{h} (t) \, \dot{S} (t) \; .
\eeq
Thus to first order in the fluctuations one has
\beq
{ 1 \over \Box^{(0)} } \cdot \Box^{(1)} (h) \cdot \left ( \, \delta G \, \bar{\rho} \, \right )=
{ 1 \over - \partial_t^2 - 3 \, {\dot{a} \over a} \, \partial_t } \cdot
\half \, \dot{h} \left ( 3 \, {\dot{a} \over a} \, \delta G - \dot{\delta G} \right ) \bar{\rho}
\label{eq:box_ratio}
\eeq
and  there is no change from the result quoted in \cite{ht10}.
There we set $ s = 0 $, since we were only interested in 
cosmological density perturbations $ \delta $, 
which couple only to the trace part of the gravitational field fluctuations $ h_{ij} $.

\vfill

\newpage


\newpage


\begin{thebibliography}{99}


\bibitem{dam06}
T.~Damour,  
% {\sl Experimental tests of Gravitational Theory},
in Review of Particle Physics, J.\ Phys.\ {\bf G} 33, 1 (2006);
update in
http://pdg.lbl.gov/2009/reviews/rpp2009-rev-gravity-tests.pdf
(Nov. 2009).

\bibitem{dam93}
T.~Damour and K.~Nordtvedt,
  %``General relativity as a cosmological attractor of tensor scalar theories,''
  Phys.\ Rev.\ Lett.\  {\bf 70}, 2217 (1993); \\
T.~Damour and G.~Esposito-Farese,
%``Testing gravity to second postNewtonian order: A Field theory approach,''
  Phys.\ Rev.\  D {\bf 53}, 5541 (1996).
  
\bibitem{ven02}
G.~Veneziano,
 ``String cosmology: The pre-big bang scenario,''
  hep-th/0002094; \\
% \bibitem{gas07}  
M.~Gasperini and G.~Veneziano,
% ``String Theory and Pre-big bang Cosmology,''
 hep-th/0703055.

\bibitem{wil72}
K.~G.~Wilson,
 %``Feynman graph expansion for critical exponents,''
Phys.\ Rev.\ Lett.\  {\bf 28}, 548 (1972);  
 %``Quantum field theory models in less than four-dimensions,''
Phys.\ Rev.\ D {\bf 7}, 2911 (1973); see also: \\
%\bibitem{par76}
G.~Parisi, in the proceedings of the 1976 Carg\`ese Summer Institute, vol. 26
(Plenum Press, New York 1977); \\
%\bibitem{bre76}
E.~Brezin and J.~Zinn-Justin,
  %``Renormalization of the nonlinear sigma model in 
  % 2 + epsilon dimensions. Application to the Heisenberg ferromagnets,''
 Phys.\ Rev.\ Lett.\ {\bf 36}, 691 (1976);
  Phys.\ Rev.\  {\bf D14}, 2615 (1976).

\bibitem{wei77}
S.~Weinberg, in {\sl `General Relativity - An Einstein Centenary
Survey'}, edited by S.W. Hawking and W. Israel,
(Cambridge University Press, 1979).

\bibitem{zin02}
J.~Zinn-Justin,
{\sl Quantum Field Theory and Critical Phenomena},
Oxford University Press (New York, fourth edition, 2002).

\bibitem{hbook}
H.~W.~Hamber, {\sl Quantum Gravitation}, 
(Springer Publishing, Berlin and New York, 2009), and references therein.

\bibitem{hw84}
H.~W.~Hamber and R.~M.~Williams, 
  %``Higher Derivative Quantum Gravity On A Simplicial Lattice,''
  Nucl.\ Phys.\  B {\bf 248}, 392 (1984); 
 ibid.  {\bf 260}, 747 (1985);
  %``Simplicial Quantum Gravity With Higher Derivative Terms: 
  %  Formalism And Numerical Results In Four-Dimensions,''
  ibid. {\bf 269}, 712 (1986);
  %``Nonperturbative Simplicial Quantum Gravity,''
  Phys.\ Lett.\  B {\bf 157}, 368 (1985);
 %``Newtonian potential in quantum Regge gravity,''
Nucl.\ Phys.\ {\bf B435} 361 (1995).
 % On the Measure in Simplicial Quantum Gravity
 %Phys.\ Rev. D {\bf 59}, 064014 (1999).

\bibitem {ham00}
H.~W.~Hamber,
Phys.\ Rev.\ {\bf D 45}, 507 (1992);
% {\it Phases of Simplicial Quantum Gravity in Four Dimensions:
% Estimates for the Critical Exponents}, 
Nucl.\ Phys. {\bf B400}, 347 (1993);
Phys.\ Rev.\ {\bf D 61}, 124008 (2000).
\bibitem {eps}
H.~Kawai and M.~Ninomiya,  Nucl.\ Phys.\ {\bf B336}, 115 (1990); \\
H.~Kawai, Y.~Kitazawa and M.~Ninomiya,  Nucl.\ Phys.\ {\bf B393}, 280 (1993) 
and {\bf B404} 684 (1993); \\
Y.~Kitazawa and M.~Ninomiya, Phys.\ Rev.\ {\bf D55}, 2076 (1997); \\
% \bibitem{eps1}
T.~Aida and Y.~Kitazawa, Nucl.\ Phys.\ {\bf B491}, 427 (1997).

\bibitem{reu98}
M.~Reuter,
  %``Nonperturbative Evolution Equation for Quantum Gravity,''
  Phys.\ Rev.\  D {\bf 57}, 971 (1998);  \\
% \bibitem{reu08}
M.~Reuter, and H.~Weyer, 
% in {\sl Quantum Gravity: Challenges and Perspectives},  
% Bad Honneff (Hermann Nicolai ed.), 
Gen.\ Relativ.\ Gravit.\  {\bf 41}, 983 (2009); \\
% \bibitem{reu10}
  E.~Manrique, M.~Reuter and F.~Saueressig,
  %``Bimetric Renormalization Group Flows in Quantum Einstein Gravity,''
 hep-th/1006.0099, and references therein.

\bibitem {vil84}
G.~A.~Vilkovisky, in {\sl Quantum Theory of Gravity}, edited by 
S. Christensen (Hilger, Bristol, 1984);
Nucl.\ Phys.\ {\bf B234}, 125 (1984).

\bibitem {ven90}
T.~R.~Taylor and G.~Veneziano,
% Quantum gravity at large distances and the cosmological constant 
Nucl.\ Phys.\ {\bf B345}, 210 (1990);
 % Quenching the cosmological constant 
Phys.\ Lett. {\bf B228}, 311 (1989).

\bibitem{loo07}
H.~W.~Hamber and R.~M.~Williams,
% Quantum Gravity in Large Dimensions
Phys.\ Rev.\ D {\bf 73} 044031 (2006);
% Gravitational Wilson Loop and Large Scale Curvature
ibid.  D {\bf 76} 084008 (2007);
% \bibitem {loo10}
% H.~W.~Hamber and R.~M.~Williams,
% Wilson Loop in Discrete Gravity
ibid. D {\bf 81} 084048 (2010).

\bibitem {hw05}
H.~W.~Hamber and R.~M.~Williams,
% Nonlocal Effective Gravitational Field Equations and the Running
% of Newton's Constant G
Phys.\ Rev.\ {\bf D 72}, 044026-1-16 (2005).

\bibitem{hw06}
H.~W.~Hamber and R.~M.~Williams,
% Constraints on Gravitational Scaling Dimensions from
% Non-Local Effective Field Equations
  Phys.\ Lett.\ {\bf B643}, 228 (2006);
%``Renormalization group running of Newton's G: The static isotropic case,''
  Phys.\ Rev.\  D {\bf 75}, 084014 (2007).

\bibitem{lop07}
  D.~Lopez Nacir and F.~D.~Mazzitelli,
  %``Running of Newton's constant and non integer 
  % powers of the d'Alembertian,''
  Phys.\ Rev.\  D {\bf 75}, 024003 (2007).

\bibitem{ht10}
H.~Hamber and R.~Toriumi,
Phys.\ Rev.\  D {\bf 82}, 043518 (2010).

\bibitem{mab95}
C.~-P.~Ma and E.~Bertschinger,
% "Cosmological Perturbation Theory in the Synchronous and Conformal Newtonian Gauges"
{\it Astrophysical Journal}, v.455, p.7 (1995).


\bibitem {cor94}
H.~W.~Hamber,
% {\it Invariant Correlations in Simplicial Gravity},
Phys.\ Rev. {\bf D50} 3932 (1994).

\bibitem{ric79}
  J.~L.~Richardson,
  % ``The Heavy Quark Potential And The Upsilon, J / Psi Systems,''
  Phys.\ Lett.\ B {\bf 82}, 272 (1979).

\bibitem{eic81}
  E.~Eichten and F.~Feinberg,
  %  ``Spin Dependent Forces In QCD,''
  Phys.\ Rev.\ D {\bf 23}, 2724 (1981);
  W.~Buchmuller and S.~H.~H.~Tye,
  % ``Quarkonia And Quantum Chromodynamics,''
  Phys.\ Rev.\ D {\bf 24}, 132 (1981);
  U.~Ellwanger, M.~Hirsch and A.~Weber,
  % ``The heavy quark potential from Wilson's exact renormalization group,''
  Eur.\ Phys.\ J.\ C {\bf 1}, 563 (1998).

\bibitem{ben99}
  M.~Beneke,
  % ``Renormalons,''
  Phys.\ Rept.\  {\bf 317}, 1 (1999);
  M.~Beneke and V.~M.~Braun,
  % ``Power corrections and renormalons in Drell-Yan production,''
  Nucl.\ Phys.\ B {\bf 454}, 253 (1995).

% Classical GR density pert theory, growth exponent

\bibitem{pee93}
P.~J.~E.~Peebles, {\it Principles of Physical Cosmology}, (Princeton University Press, 1993).

%\bibitem{wei72}
%S.~Weinberg, 
%{\it Gravitation and Cosmology: Principles and Applications of the General Theory
%of Relativity}, (J. Wiley, 1972).

% \bibitem{wei08}
% S.~Weinberg, 
% {\it Cosmology}, (Oxford University Press, 2008).

 
% Recent constraints on gamma 0.55+-0.12
  
\bibitem{smi09}
  F.~Schmidt, A.~Vikhlinin and W.~Hu,
  %``Cluster Constraints on f(R) Gravity,''
  Phys.\ Rev.\  D {\bf 80}, 083505 (2009).

\bibitem{vik09}
  A.~Vikhlinin {\it et al.},
  %``X-ray Cluster Cosmology,''
  arXiv:0903.5320 [astro-ph.CO];
  %``Cosmological Studies With A Large-Area X-ray Telescope,''
  arXiv:0903.2297 [astro-ph.CO].

\bibitem{rap09}
D.~Rapetti, S.~W.~Allen, A.~Mantz and H.~Ebeling,
  %``The Observed Growth of Massive Galaxy Clusters III: Testing General
  % Relativity on Cosmological Scales,''
 arXiv:0909.3098; arXiv:0909.3099;  arXiv:0911.1787 [astro-ph.CO].

\bibitem{uza02}
  J.~P.~Uzan,
  %``The fundamental constants and their variation: Observational status and
  % theoretical motivations,''
  Rev.\ Mod.\ Phys.\  {\bf 75}, 403 (2003).

\bibitem{uza09} 
  J.~P.~Uzan,
  %``Tests of General Relativity on Astrophysical Scales,''
  arXiv:0908.2243 [astro-ph.CO].


% Slip function

\bibitem{ame07}
  L.~Amendola, M.~Kunz and D.~Sapone,
  %``Measuring the dark side (with weak lensing),''
  JCAP {\bf 0804}, 013 (2008).

% Paolo Serra et al 

\bibitem{dan09}
  S.~F.~Daniel {\it et al.},
  %``A Multi-Parameter Investigation of Gravitational Slip,''
  Phys.\ Rev.\  D {\bf 80}, 023532 (2009).


\end{thebibliography}
\end{document}